\documentclass[aps,pre,twocolumn,superscriptaddress]{revtex4} 
\usepackage{bm}
\usepackage{graphicx}
\usepackage{amssymb}
\usepackage{amsmath}

\newcommand{\be}{\begin{equation}}
\newcommand{\ee}{\end{equation}}
\newcommand{\bea}{\begin{eqnarray}}
\newcommand{\eea}{\end{eqnarray}}

\newcommand{\br}{\mathbf{r}}                 
                   
\newcommand{\bv}{\mathbf{v}}
\newcommand{\rd}{\mathrm{d}} 
\newcommand{\bz}{\mathbf{z}}
\newcommand{\bG}{\mathbf{G}}

\newcommand{\bE}{\mathbf{E}}
\newcommand{\bR}{\mathbf{R}}

\newcommand{\tr}{\tilde{\mathbf r}}
\newcommand{\tN}{\tilde N}
\newcommand{\tU}{\tilde U}
\newcommand{\tbm}{\tilde \mu_0}

\newcommand{\bmu}{\mbox{\boldmath${\mu}$}}           
\newcommand{\bxi}{\mbox{\boldmath${\xi}$}}   
\newcommand{\bsig}{\mbox{\boldmath${\sigma}$}}   
\newcommand{\bom}{\mbox{\boldmath${\omega}$}}

\begin{document}

\title{Hydrodynamic effects in driven soft matter}

\author{Manoel Manghi}
\affiliation{Laboratoire de Physique Th\'eorique, UMR CNRS 5152, IRSAMC, Universit\'e Paul Sabatier, 31062 Toulouse, France}
\author{Xaver Schlagberger}
\affiliation{Physik Department, Technical University Munich, 85748 Garching, Germany}
\author{Yong-Woon Kim}
\affiliation{Materials Research Laboratory, University of California at Santa Barbara, Santa Barbara, CA 93106, USA}
\author{Roland R. Netz}
\affiliation{Physik Department, Technical University Munich, 85748 Garching, Germany}

\date{\today}

\begin{abstract}
Recent theoretical works exploring the hydrodynamics of soft material in non-equilibrium situations are reviewed. We discuss the role of hydrodynamic interactions for three different systems: \textit{i)} the deformation and orientation of sedimenting semiflexible polymers, \textit{ii)} the propulsion and force-rectification with a nano-machine realized by a rotating elastic rod, and \textit{iii)} the deformation of a brush made of  grafted semiflexible polymers in shear flows. In all these examples deformable polymers are subject to various hydrodynamic flows and hydrodynamic interactions. Perfect stiff nano-cylinders are known to show no orientational effects as they sediment through a viscous fluid, but it is the coupling between elasticity  and hydrodynamic torques that leads to an orientation perpendicular to the direction of sedimentation. Likewise, a rotating stiff rod does not lead to a net propulsion in the Stokes limit, but if bending is allowed an effective thrust develops whose strength and direction is independent of the sense of rotation and thus acts as a rectification device. Lastly, surface-anchored polymers are deformed by shear flows, which modifies the effective hydrodynamic boundary condition in a non-linear fashion. All these results are obtained with hydrodynamic Brownian dynamics simulation techniques, as appropriate for dilute systems. Scaling analyses are presented when possible. The common theme is the interaction between elasticity of soft matter and hydrodynamics, which can lead to qualitatively new effects.
\end{abstract}

\maketitle

\section{Introduction}
\label{intro}

Dynamics in soft matter systems far from equilibrium is of central interest in various interdisciplinary fields like biophysics or microfluidics. Most generally, soft objects subject to thermal noise
show thermally activated shape fluctuations and are sensitive to even minute external forces.
Moreover, the dynamics of small objects evolving in liquid solvents, \textit{e.g.} water,
is modified by the presence of flow and viscosity. 
As an example, external forces exerted on micro-objects lead to long-ranged effects
due to momentum diffusion and modify  any motion and flow. 
In particular, different moving objects or different moving parts of a macromolecules
are coupled among each other \textit{via} the hydrodynamic flow fields emanating from the different regions.
Such  hydrodynamic interactions, mediated by the viscous solvent, are well studied in the context of colloids~\cite{happel,lyklema,hunter,russel} and polymers~\cite{doi}. 
They are responsible for many striking features such as propulsion at the micron scale or measurable deviations from pure Brownian random dynamics in sedimentation and electrophoresis experiments~\cite{doi,pgg,viovy}. 
For small objects and rather modest flow velocities, the non-linear term in the Navier-Stokes
equation is negligible and one obtains the Stokes equation which describes the
so-called creeping flows.
Since the hydrodynamic equations governing the flow become then linear,
so are  all response relations between forces and torques applied on rigid micro-objects and the resulting velocities~\cite{happel}. 
This linear relationship is completely defined by the shape of the objects and thus illustrates the importance of shape design in microfluidics, for instance. 
However, the issue becomes more complicated for deformable soft objects like membranes, 
soft colloids or elastic rods and macromolecules (cylindrical viruses,
synthetic or bio-polymers). 
In that case, the deformation of the object depends on the flow which thus changes hydrodynamic boundary conditions
in an intricate and non-linear fashion, since the object shape acts
back on the flow field \textit{via} hydrodynamic screening.
Solving the problem becomes much more complicated, both at the numerical and analytical
level. Some analytical results have been obtained either close to equilibrium where linear response theory applies or by invoking the  \textit{pre-averaging} approximation,
as used in the context of the Zimm-model for  dynamics of polymers~\cite{doi}. 
However, to unravel the full coupling between shape and flow in situations far from
equilibrium, simulation techniques prove very useful. 

In this review, we focus on three situations involving hydrodynamics of soft matter  
far from equilibrium: \textit{i)}~the sedimentation and orientation of elastic rods, \textit{ ii)}~the propulsion and force-rectification by a rotating semi-flexible filament, and \textit{iii)}~the deformation of a brush of grafted semiflexible polymers 
 in shear flows.
As will be discussed, the selection of these topics is 
closely related to novel emerging experimental techniques and activities. 

Example \textit{i)} is related to the effect of \textit{anomalous birefringence},
which is observed in experiments involving charged rod-like particles such
as Tobacco-Mosaic~\cite{konski}, FD viruses~\cite{kramer} or 
synthetic polyelectrolytes~\cite{Opper,Opper2} and cylindrical micelles~\cite{Hoffmann}
in electric fields. Normally, charged objects in electric fields orient parallel to 
an externally applied electric field due to the anisotropic polarizability
which is mostly caused by the easily displaceable counterion cloud. 
This is called normal birefringence. For almost all rod-like charged systems,
one finds in a certain parameter range also anomalous birefringence,
where particles orient perpendicular to the direction of the electric field
and motion. 
This effect is obtained for long particles, law salt concentration or particle concentrations beyond mutual overlap. A similar effect, namely perpendicular orientation of 
moving cylindrical objects, is in principle also observable with sedimenting
rod-like particles~\cite{Buiten}.
This phenomenon is, at present, not completely understood~\cite{Cates}.
A possible relation with a mechanism for 
hydrodynamic orientation associated with rod deformation was recently suggested~\cite{xaver}.
As follows from the linear-Stokes equation,
perfectly straight and stiff nano-cylinders  show no orientational effects
as they sediment through a viscous fluid. But an elastic rod is
deformed due to hydrodynamic forces, and the coupling between
elasticity  and hydrodynamic torques leads strikingly to an orientation
perpendicular to the direction of sedimentation.

In example \textit{ii)} the
question of propulsion of a micrometer sized object through a viscous fluid 
is addressed from an engineering point of view. Due to the linearity of
hydrodynamic equations, propulsion on the nano-meter scale
calls for design strategies very different from the macroscopic world, since
inertia plays no role and friction is the only way of 
producing thrust~\cite{purcell,purcell2}. 
The studied  propulsion mechanism is inspired by biological systems such as bacteria,
 which move by rotating propeller-like appendices, so-called flagella~\cite{lighthill,lighthill2,lighthill3}. Those propellers involve helical stiff polymers that are rotated at their base by a 
 rotary molecular motor. Hydrodynamic friction converts the rotational motion of the helix into thrust along the helix axis~\cite{lighthill,purcell}. 
 For a number of biomedical applications, \textit{e.g.} for directed motion of artificial viruses through cells or nano-devices through the bloodstream,  it is desirable to develop similar synthetic propulsion mechanisms or to incorporate biological single-molecule motors into synthetic environments. 
A second possible field of application involves mixing strategies in nano-fluidic devices,
which could be achieved by moving or rotating surface-anchored polymers~\cite{mixing}.
 Recent discoveries opened the route to the synthetic manufacture of rotary single-molecule motors driven by chemical~\cite{kelly} or optical~\cite{koumura,harada} energy. An ATPase can also be fixed to different substrates and used to rotate metallic~\cite{Soong} or organic nano-rods~\cite{Noji}. All these experiments raise the question about the minimal design necessary to convert the rotational power of such nano-engines into directed thrust in a viscous environment. Since stiff helices are actually difficult to manufacture down to the micron scale, we investigate
whether  a  straight \textit{elastic} filament can act as a propeller. 
Our  design consists of a straight elastic rod that rotates around a point with a constrained azimuthal angle, \textit{i.e.}, on the surface of a cone. In the absence of elasticity, no effective thrust is produced
as one averages the motion over one whole cycle of rotation. But a rod with finite bending
modulus will be deformed and in fact take on a shape that resembles one period of a 
helix. Due to the symmetry breaking, a net thrust is obtained which in fact does not 
depend on the sense of rotation of the rod: the device acts as a nano force-rectification 
device~\cite{manoel}.

Finally, we study the influence of a shear flow on the conformations of surface-anchored
semiflexible polymers. When the shear rate increases, polymers will be oriented
in the direction of the shear and be bent down in order to be more and more parallel to the surface. These conformational effects clearly depend also on the grafting density of polymers and on the stiffness of polymers. As polymers bend down onto the surface, 
the hydrodynamic boundary condition changes and the flow can penetrate closer
towards the surface. Again, it is the coupling between hydrodynamic flow effects
and elastic deformation of polymers which leads to intricate non-linear effects.
These phenomena have been experimentally investigated using colloidal spheres
with surface anchored DNA molecules, which were being held fixed using a laser trap
in uniform flows of different strengths~\cite{Gutsche}. Since the force acting
on the sphere is measured in such an experiment, the hydrodynamic drag and therefore
the position of the surface of shear can be measured with high precision. 
Similar phenomena also appear in the glycocalix layer, which is a polymer-brush like
coating of the endothelial cell layers in blood vessels. The shear rates in 
blood stream can be enormous, and conformational changes of the glycocalix layer
has important consequences on permeability of drugs, nutritional agents or viruses.

The paper is divided in five sections. In Section~\ref{nano}, we give an introduction of hydrodynamics at low Reynolds numbers and focus particularly on hydrodynamic interactions between two moving spheres in a viscous fluid. 
The integral representation of the flow velocity, introduced  by Oseen using 
Green's function techniques, is developed, since it is at the heart of our simulation methods. 
We present Green's tensors both for an unbounded fluid and a semi-infinite fluid medium with a no-slip wall. 
Section~\ref{non_eq} presents our model of semi-flexible polymers under external force  which
is  implemented in Brownian dynamics simulations in the presence of hydrodynamic interactions. 
Section~\ref{soft} is devoted to hydrodynamics of elastic polymers under external force load. 
After recalling some basic results for rigid cylinders and helices (Section~\ref{rigid}), 
we focus on the coupling between hydrodynamic interactions and elastic deformations.
This coupling leads to the orientation of elastic cylinders under an homogeneous field,
 propulsion with rotating polymers, and deformation of brush layers in shear flows.

\section{Hydrodynamics on the nanoscale}
\label{nano}

\subsection{Hydrodynamics at low Reynolds number}
\label{hydro}

The Navier-Stokes equation combines both convective and viscous terms, the ratio of which is given by the Reynolds number, $Re=\frac{\rho U L}{\eta}$, where $U$ and $L$ are the typical velocity and size of the flow, $\rho$ the mass per unit volume and $\eta$ the shear viscosity of the fluid~\cite{landau}. In this paper, we are dealing with small objects moving slowly in a fluid, like a polymer, a macromolecule or a latex bead. The typical scales are in the range $L\simeq 10^{-8}$--$10^{-5}$m and $U\simeq 10^{-8}$--$10^{-4}$m/s yielding in water ($\rho=10^3$kg/m$^3$, $\eta=10^{-3}$Pa/s) $Re\simeq 10^{-10}$--$10^{-3}$. Hence, hydrodynamics on the nanoscale
belong to the small Reynolds number regime, $Re\ll 1$. In this framework, viscous damping (momentum diffusion) dominates inertia (momentum convection), and the creeping flow produced by moving objects is described by the Stokes equations~\cite{happel,landau}
\bea
\eta \nabla^2\mathbf{v}(\br)-\nabla p(\br)+\mathbf{f}\delta(\br) &=& 0 \label{stokes}\\
\nabla\cdot\mathbf{v}(\br) &=& 0 \label{incompress}
\eea
where $p(\br)$ is the pressure field, $\mathbf{v}(\br)$ the fluid velocity and $\mathbf{f}$ a point force at $\br=0$. In this paper, the fluid can be taken as incompressible at the time scales we consider.

The solution of the linear equations~(\ref{stokes})-(\ref{incompress}) can be 
found using Green function techniques~\cite{lad,kim2,jackson}.
The Oseen tensor~\cite{oseen}, also called Stokeslet, $\mathbf{G}(\br)$ describing the fluid velocity disturbance caused by a point force exerted at the origin, is
\bea
\mathbf{G}(\br) &=& \frac{\mathbf{1}}{4\pi \eta r}-\nabla\otimes\nabla\frac{r}{8\pi\eta} \nonumber \\ &=& \frac{1}{8\pi\eta r}\left(\mathbf{1}+\frac{\br\otimes\br}{r^2}\right)  \label{oseen}\\
p(\br) &=& -\nabla\cdot\left(\frac{\mathbf{f}}{4\pi r}\right)
\label{pressure}
\eea
where $r=|\br|$ and $\br\otimes\br=\br \br^{\mathrm{T}}$ is a dyadic tensor. 
The pressure field is the potential of a dipole distribution of strength $-\mathbf{f}$. 
Since we are dealing with an unbounded fluid, the flow disturbance vanishes at infinity: $v(r\to\infty)=0$. The famous inverse first power dependence in $r$ means that hydrodynamic interactions are \textit{long-ranged}. Moreover the flow velocity $\mathbf{v}(\br)$ is not uniform, the perturbation being larger in the direction of $\mathbf{f}$. This non-uniformity is linked  to the symmetry-breaking into longitudinal and transverse modes in fluids. 

The flow around any moving rigid body (volume $V$ and surface $S$) in an unbounded medium can then be theoretically constructed by using a distribution of Stokeslets on the body surface. Indeed, if there is no source of momentum in the fluid, the flow is only determined by the no-slip condition at the body surface. For a steady creeping flow, the propulsive external force applied to a moving object is balanced by the frictional force exerted by the fluid on the body surface, which is given by integrating the stress tensor
\be
\bsig=-p\mathbf{1}+\eta [\nabla \otimes \bv  + (\nabla \otimes \bv)^T]
\label{stresstensor}
\ee
over the body surface~\cite{landau}. 
Most generally, the bulk fluid velocity is then given by the surface integral
\be
\mathbf{v}(\br)=\int_{S} \mathbf{G}(\br-\br')\cdot \mathbf{f}(\br')\mathrm{d}^2\br'
\label{convol}
\ee
where the surface force distribution $\mathbf{f}(\br')$ is the source of the flow field.

As an illustration of these concepts, we calculate in the following the flow induced by a moving sphere in an unbounded fluid. It will allow us to introduce hydrodynamic interactions between spherical particles evolving in a creeping flow.

\subsection{Flow around a moving sphere and hydrodynamic interactions}
\label{sphere}

For a rigid object with an unspecified shape, the calculation of the flow field at distance $r$ from the immersed object of size $a$ using Eq.~(\ref{convol}) becomes hardly tractable and, similarly to electrostatics, a multipole expansion can be done in powers of $a/r$. At first order, one can approximate the velocity  distribution by a point force exerted at the centre of the body. This corresponds to the far field solution, which is a Stokeslet. However, this approximation is no longer valid when we consider the flow field closer to the immersed body. We shall then take into account its finite size. Let us consider a rigid sphere of radius $a$ with its centre positioned at $\br_i$. 
We  rewrite Eq.~(\ref{convol}) using a Taylor expansion and obtain the fluid velocity 
$\bv_i(\br)$, associated to a uniform force density $\mathbf{f}_i ({\bf r})=\mathbf{f}_i$ applied on the sphere surface
\be
\bv_i (\br) = \left( 1+\frac{a^2}{6}\nabla^2_{\br_i}\right) \mathbf{G}(\br-\br_i)\cdot\mathbf{F}_i \label{doublet}
\ee
where $\mathbf{F}_i= \oint_S\mathbf{f}_i(\br')\rd^2 \br' = 4\pi a^2 \mathbf{f}_i$ is the total force acting on the particle $j$. This result is indeed exact for a sphere since its high symmetry ensures that higher order surface integrals are zero. 

One can now calculate the velocity of the rigid sphere, $\bv_i^s$, by utilizing the \textit{no-slip boundary condition} on the sphere surface: the sphere velocity is 
 equal to the fluid velocity at the surface. We thus find the Stokes law for the velocity of a sphere which moves slowly under a force $\mathbf{F}_i$ as~\cite{happel,lamb}
\be
\bv_i^s = \frac1{4\pi a^2}\oint_S \bv_i(\br) \,\rd^2\mathbf{r}  = \frac{\mathbf{F}_i}{6\pi \eta a} 
\label{Stokes formula}
\ee
Since the sphere is moving with a stationary velocity, $-\mathbf{F}_i$ is also the frictional force acting on a sphere moving at velocity $\bv^s_i$.

Suppose now that we have two spheres $i$ and $j$ of same radius $a$ in an unbounded fluid with two external forces $\mathbf{F}_i$ and $\mathbf{F}_j$ acting on them. The fluid velocity generated by $\mathbf{F}_j$ on particle $j$ also tends to drive the sphere $i$: these interactions \textit{via} momentum diffusion are called \textit{hydrodynamic interactions}. Of course, the modified velocity of particle $i$ should in return modify the flow field close to particle $j$ and then its velocity and so on. However, these effects are successively smaller by a factor $2a/ r_{ij}$ where $\br_{ij}=\br_i-\br_j$ and can be calculated by the method of reflections~\cite{happel,felderhof}. In the following, we consider that the spheres are widely separated, $2a/ r_{ij}\ll 1$, and the velocity of particle $i$ is simply the sum of two terms: its self-propulsion by $\mathbf{F}_i$ given by Eq.~(\ref{Stokes formula}) and the contribution of the flow created by the sphere $j$, $\bv_j(\br)$, given by Eq.~(\ref{doublet}). The sum of the two effects can be written as
\be
\label{HI}
\bv^s_i =  \frac{\mathbf{F}_i}{6\pi \eta a} + \frac1{4\pi a^2}\oint_{S_i}\bv_j(\br')\, \rd^2\mathbf{r'} = \sum_j \bmu_{ij}\cdot \mathbf{F}_j 
\ee
In the above equation we use the Taylor expansion of the flow field 
with respect to the particle centre $\br_i$. This naturally leads to the mobility tensor, 
that describes hydrodynamic interactions between particles, as
\bea
\bmu_{ii} &=& \mu_0 \mathbf{1} = \frac{1}{6\pi\eta a}\mathbf{1}\\
\bmu_{ij}(\br_{ij}) &=& \left( 1+\frac{a^2}{6}\nabla^2_{\br_{ij}}\right)
 \left( 1+\frac{a^2}{6}\nabla^2_{\br_{ij}}\right) \bG(\br_{ij}) \nonumber\\
		     &=&  \frac{1}{8\pi\eta\, r_{ij}}\left(\mathbf{1} + \frac{\br_{ij} \otimes \br_{ij}}{r_{ij}^2} \right) \nonumber \\
		     &+& \frac{a^2}{4\pi\eta\, r_{ij}^3}\left(\frac{\mathbf{1}}{3} - \frac{\br_{ij} \otimes \br_{ij}}{r_{ij}^2} \right)
\label{RP2}
\eea
where $\mathbf{1}$ is the $3\times3$ unit matrix. One notes that this mobility tensor includes the lowest corrections of particle size over the Oseen tensor description, and corresponds to the well-known Rotne-Prager tensor in an unbounded fluid~\cite{RP}. One thus should keep in mind that our treatment of hydrodynamic interactions is only approximate for particles which are separated by a distance on the order of $a$.

\subsection{Hydrodynamic interactions near a no-slip surface}
\label{surf}

\begin{figure}[ht]
\begin{center}
\includegraphics[height=8cm]{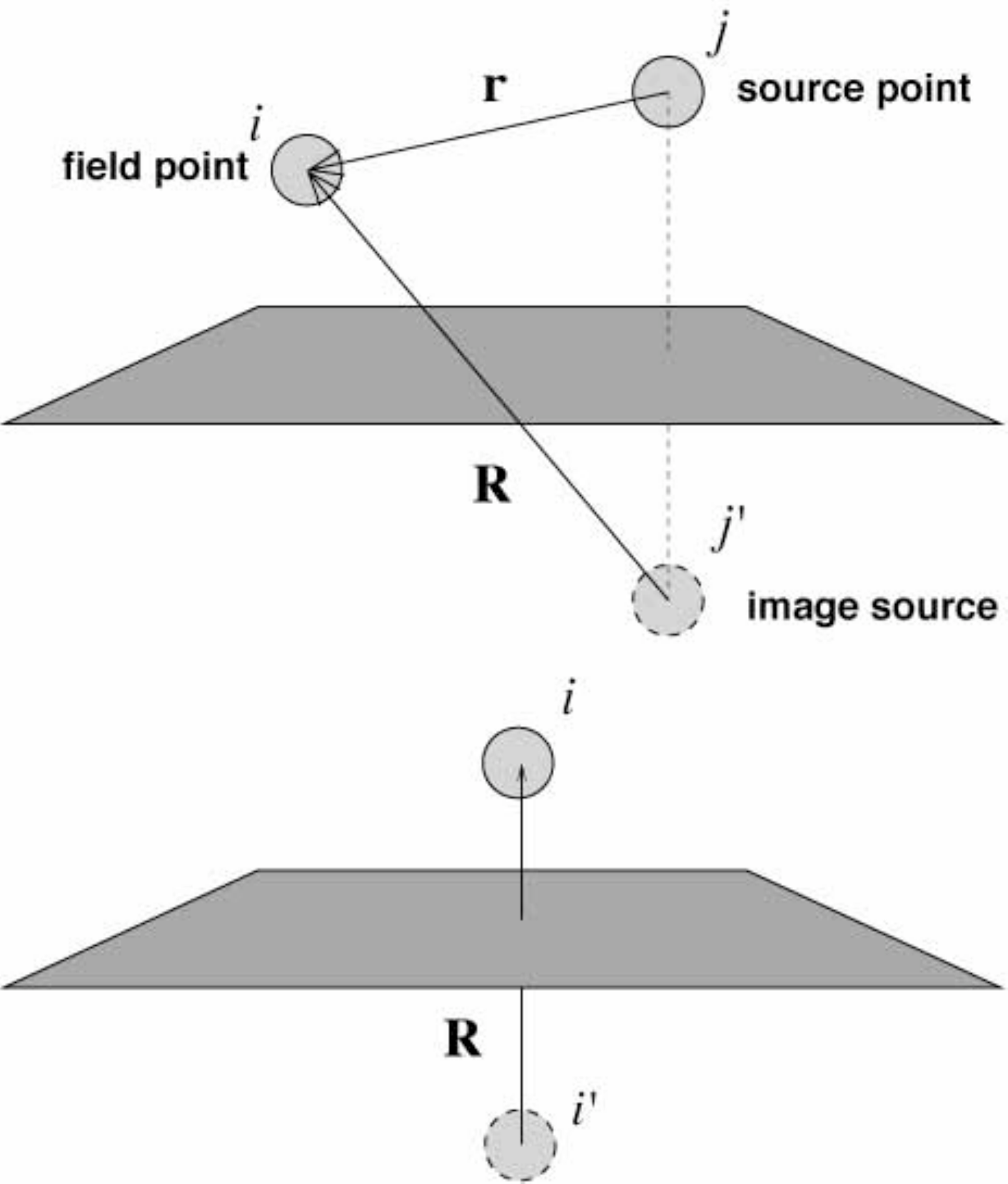}
\caption{Top: Two spherical particles $i$ and $j$, where $\br=\br_i-\br_j$, in the presence of no-slip planar surface. The dashed sphere $j'$ beneath the surface represents the image of $j$-th particle and $\bR=\br_i-\br_{j'}$. Bottom: Schematic picture for determining self-mobility of a single particle where $\bR$ is the relative position vector between the particle and its own image.}
\label{fig:scheme}
\end{center}
\end{figure}

If a particle is moving near a rigid boundary like a solid planar surface, the situation becomes much more complicated due to boundary conditions. The fluid velocity must vanish at the surface which is at rest. Here again, theoretical methods used in electrostatics apply. Blake~\cite{blake} developed the image method for hydrodynamics near a plane and obtained the Green's function in the presence of a 
no-slip planar boundary at $z=0$ as
\be
\label{Eq:wall}
\bG_{wall} (\br_i,\br_j) = \bG(\br) - \bG(\mathbf{R}) + \bG^D(\mathbf{R}) - \bG^{SD}(\mathbf{R})
\ee
where $\br=\br_i-\br_j$ and $\mathbf{R} = \br_i-\br_{j'}$ and $j'$ is the image with respect to the plane 
$z=0$ (see Fig.~\ref{fig:scheme}). The Blake's Green function relates the flow at $\br_i$ to a unit point force applied at $\br_j$ in the \textit{bounded} fluid medium. One notes that the no-slip boundary condition on the plane of the wall is satisfied by the image system~\cite{jones,jones1,jones2,brenner1,brenner2,onell,jeff}, consisting of the original Stokeslet $\bG(\br)$, the image Stokeslet $\bG(\bR)$, a Doublet $\bG^D(\bR)$ and a source Doublet $\bG^{SD}(\bR)$. If one uses the tensor $\bG_{wall}$ 
given by Eq.~(\ref{Eq:wall}), instead of $\bG^S$ given by Eq.~(\ref{oseen}), in equation~(\ref{RP2}), we obtain the equivalent of the Rotne-Prager mobility tensor in the presence of a no-slip wall,
which is the level of description we chose for the last example of surface-anchored polymers in shear
flows.

\section{Hydrodynamic non-equilibrium simulations of soft matter}
\label{non_eq}

\subsection{Review of numerical methods}

On the methodological side, the adequate inclusion of hydrodynamics in a theoretical description of soft-matter systems has been a long-standing problem~\cite{allen,brady,nanofluidics}. From a simulation point of view, molecular dynamics (MD) simulations with explicit consideration of solute and solvent take into account hydrodynamic interactions on an adequate level. However, most of the simulation time is used to evaluate the solvent dynamics in great detail, which is not necessary to achieve the proper hydrodynamic behavior on the colloidal or polymer length scale: indeed, MD simulations require time-resolution on the order of $10^{-15}$s to describe inter-molecular forces whereas the typical diffusion time of a colloid of size $a=10$nm in water is $\tau=6\pi\eta a^3/k_BT\simeq 5\times 10^{-6}$~s. Moreover, it will displace approximately $10^8$ solvent molecules. This large length- and time-scale gap between the solvent molecules and the solute calls for a coarse-grained and simplified description of the solvent dynamics.

This has led to the development of novel mesoscopic simulation techniques. Prominent examples are lattice models such as lattice-Boltzmann methods~\cite{13,13bis,succi,14}, particle based off-lattice methods such as dissipative particle dynamics~\cite{DPD,15,16}, and multi-particle-collision dynamics~\cite{17,18}, and more coarse-grained scheme such as Lowe-Anderson thermostat~\cite{lowe,lowe1} and stochastic rotation dynamics~\cite{male}. 

We next describe the implementation of a Brownian dynamics simulation
for elastic polymers. We first  introduce the general framework  of Langevin dynamics with hydrodynamic interactions and then adapt it to semi-flexible polymers. 
In contrast to lattice-based methods, for which the numerical complexity scales with the system volume, hydrodynamic simulation methods~\cite{19,20,21} lead to a numerical complexity which scales with the number of monomers included in the simulation. Clearly, hydrodynamic simulation methods will be advantageous whenever dilute systems are being studied and the number of polymers is small. Likewise, semi-infinite systems (\textit{e.g.} polymers grafted or adsorbed to a single solid surface and connected to bulk solvent) can be straightforwardly simulated using hydrodynamic simulation methods and the appropriate hydrodynamic GreenÕs function satisfying \textit{e.g.} the no-slip boundary condition at the surface. 
Following this scheme, the solvent is treated as a continuum, which is 
acceptable for flows on length scales larger than the size of solvent molecules~\cite{nanofluidics}. 
Stokesian dynamics does not allow to describe the real flow field in the immediate vicinity of our structural units, the monomers, whose size ranges between $1-10$~nm. 
On the other hand, the flow field calculated far away form the monomers by using the Rotne-Prager tensor, Eq.~(\ref{RP2}) remains exact at the order of $(a/r)^2$ and is not influenced by the flow close to the monomers. Moreover, the hydrodynamic interactions between two monomers in contact are negligible compared to direct monomer-monomer interactions (stretching and bending) or excluded volume effects. Hence, our treatment of hydrodynamic interactions between monomers far apart should remain valid. 
This is borne out by recent Non-Equilibrium Molecular Dynamics Simulations of  sedimenting
carbon nanotubes in water, where it was demonstrated that the Stokes
equation describes simulated solvent flow-profiles down to the sub-nanometer
length scales~\cite{Walther}. More subtle is the question of the appropriate
hydrodynamic surface-boundary condition, as it depends on surface details
and displays variable amounts of slip at hydrophobic surfaces~\cite{Walther}.
In our case we are mostly concerned with hydrophilic objects, where the no-slip
boundary condition is valid to a very good degree.

\subsection{Brownian hydrodynamics}
\label{brown}

The two following assumptions provide a frame of reference for the description of Langevin dynamics
\begin{itemize}
	\item the mesoscopic size of polymers and colloids ($a\sim$ 10 nm to 1~$\mu$m) ensures the validity of Langevin dynamics. Fluid particles are then treated as variables having \textit{fast} dynamics and are integrated out, whereas material particles follow  \textit{slow} dynamics and are treated explicitly;
	\item we assume \textit{local thermodynamic equilibrium}, \textit{i.e.} particle velocities obey fast dynamics which implies that the memory of velocities is lost much quicker than the time scales of interest. We are thus interested in time scales larger than the momentum relaxation time, $t\gg \tau_m=m/(6a\eta)$, which constitutes the diffusive regime.
\end{itemize}

Our polymers are coarse-grained at the nanometer scale and modeled as an assembly of $M+1$ spherical particles submitted to external and interaction potentials. In this framework, we are able to write the time evolution of all sphere positions $\mathbf{r}_i(t)$, 
which is governed by the position-Langevin equation~\cite{lax,zwanzig,fixman,ermack1,ermack,durlofsky}
\bea
\dot{\mathbf{r}}_i(t) &=& \sum^M_{j=0} \bmu_{ij} \cdot\left[- \nabla_{\br_j} U(\{\mathbf{r}_k\}) + \mathbf{F}_j^{\mathrm{ext}}\right]\nonumber \\ 
& & + \;\;k_BT\sum^M_{j=0} \nabla_{\br_j}\cdot \bmu_{ij} + {\bm \xi}_i(t)
\label{langevin}
\eea
which relates the velocity of sphere $i$ to forces applied to all beads including
hydrodynamic interactions as given by  Eq.(\ref{HI}). One notes that the hydrodynamic 
mobility tensor $\bmu_{ij}$ introduces long-ranged interactions and
couples distant particles.
The stochastic term, the Langevin random displacement ${\bm \xi}(t)$, mimics 
the action of a thermal  heat bath and obeys the fluctuation-dissipation relation
\be
\left\langle{\bm \xi}_i(t)\otimes{\bm \xi}_j(t')\right\rangle=2k_BT\bmu_{ij}\delta(t-t')
\ee
which is numerically implemented by a Cholesky decomposition~\cite{ermack}. 
We neglect all memory effects in this work.
In the most part, we study the dynamics of elastic rods  in an unbounded fluid. 
In this case we use the mobility tensor ${\bmu}_{ij}$ given by Eq.~(\ref{RP2}) 
and it is straightforward to check that $\nabla_{\tr_j} \cdot {\bmu}_{ij} =0$ for every bead
and the gradient-mobility term in the Langevin equation vanishes identically.
This is different in the presence of a no-slip surface, where this term is non-zero 
and has to be taken into account.

The forces have two contributions: the internal forces, which are derivatives of the interaction potential $U$, and the external force, $\mathbf{F}_j^{\mathrm{ext}}$, applied on bead $j$. These forces depend on the specific details of the studied system and will be described in the following sections. Hydrodynamic interactions between spheres are implemented \textit{via} the Rotne-Prager mobility tensor given by Eq.~(\ref{RP2}) for an unbounded medium and by using the procedure defined in Section~\ref{surf} for a semi-infinite medium with a no-slip wall at $z=0$. It ensures that the mobility matrix is positive-definite which is necessary to use the Cholesky decomposition~\cite{ermack}.

For the numerical iterations, we discretize Eq.~(\ref{langevin}) with a time step $\Delta$ and rescale all lengths by the sphere radius $a$ according to $\tr_i=\br/a$. The iterative Langevin equation in terms of the discrete time variable $n=t/\Delta$ now reads
\bea
\tr_i(n+1) &=& \tr_i(n)+\sum^M_{j=0} \tilde{\bmu}_{ij} \cdot \left[-\nabla_{\tr_j}\tilde{U}(n) + \tilde{\mathbf{F}}_j^{\mathrm{ext}}(n) \right] \nonumber\\  
& & + \sum_{j=0}^M \nabla_{\tr_j} \cdot \tilde{\bmu}_{ij} + \sqrt{2\tilde{\bmu}_{ij} } \cdot \tilde{\bxi}_i(n)
\eea
where $\tilde{U}=U/k_BT$ is the dimensionless potential, $\tilde{\mathbf{F}}_j^{\mathrm{ext}}=a\mathbf{F}_j^{\mathrm{ext}}/k_BT$ the 
rescaled external force, and the rescaled random displacement has variance unity $\langle\tilde{\bxi}_i(n)\otimes\tilde{\bxi}_j(m)\rangle=\delta_{ij}\delta_{nm}$. 
We express all mobilities in dimensionless form by defining 
$\mathbf{\tilde{\bmu}}_{ij} = \Delta\bmu_{ij} k_BT/a^2$. The rescaled bare mobility
\be
\label{Eq:singlemo}
\tbm =  \frac{k_BT}{a^2} \Delta\mu_0
\ee
is the diffusion constant in an unbounded space in units of the particle radius $a$ and time step $\Delta$. 

\subsection{Semi-flexible polymer model}
\label{polym}

An elastic polymer is modeled as a chain of connected spheres (monomers) that interact \textit{via} stretching and bending forces. The non-dimensional elastic potential $\tilde{U}$ is 
 the discrete version of the {\it extensible} worm-like chain model
\be
\label{Utot}
\tilde{U}(\{\br_k\}) =  \sum_{i=0}^{M-1} \left[ \frac{\tilde{\gamma}}4
 (| \tilde{\bf r}_{i} -\tilde{\bf r}_{i+1}|  -2)^2
                            +\frac{\tilde{\varepsilon}}{2}\,(1-\cos \theta_i) \right]
\ee
where $\theta_i$ is the angle between neighbouring bonds of sphere~$i$. The parameters $\tilde{\gamma}=\gamma a/ k_BT$ and $\tilde{\varepsilon}=\varepsilon/(a k_BT)$ are the adimensionnal stretching and bending moduli. For an \textit{isotropic elastic cylinder} with radius $a$, bending and stretching moduli are determined by Young's modulus $E_Y$ as $\varepsilon = E_Y \pi a^4/4$ and $\gamma= E_Y \pi a^2$, which gives the relation $\tilde{\gamma} = 4\tilde{\varepsilon}$. If not 
stated otherwise, we always consider isotropic elastic rods in the following. The persistence length, \textit{i.e.} the correlation length of tangent vectors along the polymer contour, 
is given by $\ell_{\mathrm{p}}=\varepsilon/(k_BT)$. Twist and torsional degrees are omitted since for most synthetic polymers, free rotation around the polymer backbone is possible. The rescaled bending rigidity is given by the ratio of the persistence length and the rod length
\be
\frac{\ell_{\mathrm{p}}}{L}=\frac{\varepsilon}{2aMk_BT}
\ee
where $L=2a M$ is the polymer length. 

The external force can be caused by an external homogeneous field, electric or gravitational, 
\be
\label{tildeE}
\tilde{\mathbf{F}}_j^{\mathrm{ext}}=\tilde{\bE}=\frac{aqe\bE}{k_BT}
\ee
where $qe$ is the electric charge of the bead ($q$ is the valency) for an electric field or
the mass for a gravitational field as applicable to sedimentation experiments. 
The external force can also act on only part of the polymer, 
as for example the force applied by a molecular motor.

Mutual penetration of monomers is prevented by an excluded volume interaction modeled by a truncated Lennard-Jones interaction
\be
\tU_{LJ}=\omega \sum_{i<j} \left[\left(\frac2{\tr_i-\tr_j}\right)^{12} - 2\left(\frac2{\tr_i-\tr_j}\right)^6+1\right]
\ee
valid  for separation $|\tr_i-\tr_j|<2$ with an energy parameter $\omega=3$.

Lateral periodic boundary conditions are implemented with the Lekner-Sperb summation scheme~\cite{andre} for hydrodynamic and electrostatic interactions. 
For sufficient numerical accuracy we choose time steps in the range $\tbm = 10^{-3}$--$10^{-8}$. Output values are calculated every $10^3$--$10^4$ steps, total simulation times are of the order
of $10^8$--$10^9$ steps,  giving errorbars typically smaller than the symbol size.

\section{Hydrodynamics of rigid rods}
\label{rigid}

We have seen that one of the principal feature of small Reynolds numbers hydrodynamics is that the fluid velocity field exhibits a linear dependence on exerted forces on immersed solid objects. 
In this paragraph, we explore this feature in some detail and demonstrate that propulsion at low Reynolds numbers is intimately related to shape design. 
The  torque, assumed to act at the origin $\br=0$, and force
exerted on a body are calculated by integrating the stress tensor $\bsig$, Eq.~(\ref{stresstensor}), 
over the particle surface as~\cite{happel}
\bea
\mathbf{F} &=& \oint_S \bsig \cdot \rd\mathbf{S} \label{force}\\
\mathbf{N} &=& \oint_S \br \times \bsig \cdot \rd\mathbf{S} \label{torque}
\eea
It is plausible that the body shape -- and thus any symmetry properties of the solid --  influence the flow at short distances, $r\gtrsim a$. However for large distances $r\gg a$, we remind the reader that the flow field is invariably given by the Stokeslet, Eq.~(\ref{oseen})--(\ref{pressure}). For a rigid object, the velocity $\bv$ at any point $\br$ belonging to the solid is obtained by decomposing its motion into a global translation of velocity $\bv^s$ and a rotation with angular velocity $\bom^s$ related by $\bv=\bv^s+\bom^s \times \br$. Using the no-slip condition at the body surface, one can then obtain the linear relationship between the forces $\mathbf{F}$ and torques $\mathbf{N}$ exerted on the rigid object and 
translational as well as rotational velocities
$\bv^s$ and $\bom^s$ by inverting equations (\ref{force})--(\ref{torque})
\be
{\bv^s \choose \bom^s}  = \left( \begin{array}{cc} \bmu_{t} & \bmu_{tr} \\ \bmu_{rt} & \bmu_{r} \end{array}\right) {\mathbf{F} \choose \mathbf{N} }
\label{mobilities}
\ee
where the tensors $\bmu_t$ and $\bmu_r$ are the translation and rotation mobility tensors. Hence the off-diagonal tensors, $\bmu_{tr}$ and $\bmu_{rt}$, are the two coupling tensors and represent cross-effects where rotation is induced by forces and translation induced by torques. Using the generalized reciprocal theorem~\cite{happel}, one can prove that the whole mobility tensor defined in Eq.~(\ref{mobilities}) is symmetric. 
Moreover, if the solid object exhibits some type of symmetry, they will be reflected by symmetries of the mobility tensors $\bmu$. 

In the following we are interested in bodies with three mutual orthogonal planes of symmetry (orthotropic bodies) such as ellipsoids, spheres, cylinders or cubes. In this case, we find $\bmu_{rt} = \bmu_{tr}=0$ for every coordinate system. Hence translational and rotational motions are completely decoupled and \textit{only a torque can induce a rotation of the body}. Moreover, tensors $\bmu_t$ and $\bmu_r$ are diagonal when expressed in the principal axes coordinate frame~\cite{happel}. Hence for a cylinder, we have only two eigenvalues and the translation mobility tensor simplifies into
\be
\bmu_t = P \left( \begin{array}{ccc} \mu_{\perp} & 0 & 0 \\ 0 & \mu_{\perp} & 0\\ 0 & 0 & \mu_{||} \end{array}\right)P^T  \label{SB}
\ee
where $P$ is the rotation matrix into the principal axes coordinate frame. The two mobility constants $\mu_{||}$ and $\mu_{\perp}$ correspond to motions parallel and perpendicular to the cylinder, respectively. The exact values of these mobilities are only available for ellipsoids from which the values for
 a finite cylinder of length $L$ have been approximately deduced. The literature gives the following approximate values~\cite{happel}
\bea
\mu_{\perp} 	&=& \frac{2\ln(L/a) +1}{8\pi\eta L}\\
\mu_{||} 		&=& \frac{2\ln(L/a) -1.44}{4\pi\eta L}
\eea
The important point is that for sufficient long cylinders these values can be legitimately approximated by \be
\mu_{||} \simeq 2\mu_{\perp}
\ee

\begin{figure}[ht]
\begin{center}
\includegraphics[height=8cm]{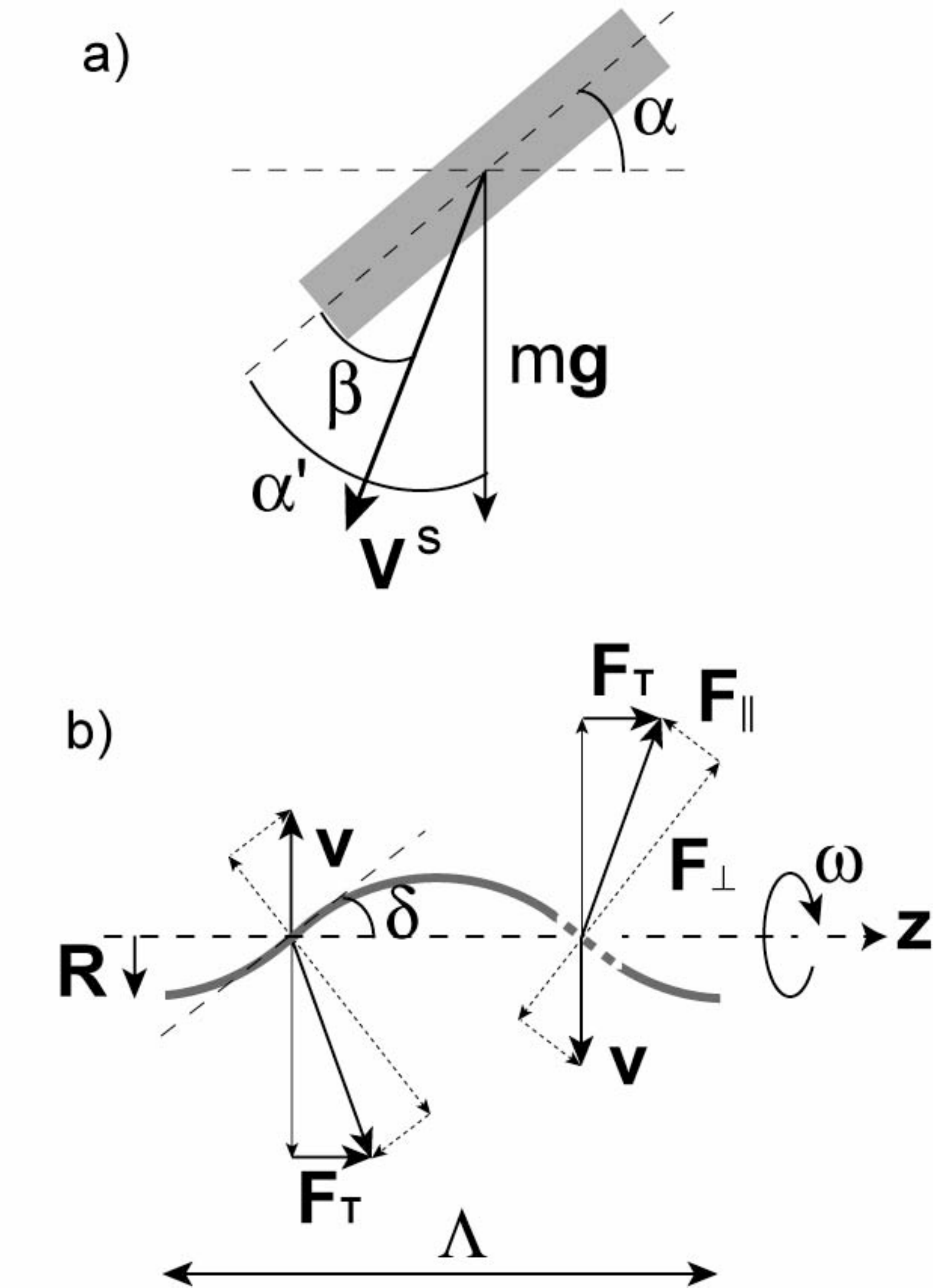} 
\caption{a) Oblique fall of a cylinder under gravity. b) Schematic analysis of 
 the viscous drag on two cylinders of a helical filament contributing to thrust. The helix is left-handed and $\bom=\omega\hat{\bz}$. The velocity of each segment is decomposed into normal and parallel velocities, $v_{\perp}$ and $v_{||}$. Hence normal and tangential friction forces, $F_{\perp}$ and $F_{||}$, act in opposite direction and their ratio is $F_{\perp}/F_{||}\simeq 2 v_{\perp}/v_{||}$. The total force has thus a component parallel to $\hat{\bz}$, $F_T$.}
\label{helixBerg}
\end{center}
\end{figure}

Suppose that a cylinder is falling under its gravitationnal weight in a quiescent fluid at low Reynolds number. The angle $\beta$ between its symmetry axis and the cylinder velocity $\bv^s$ is then given by (see Figure~\ref{helixBerg}a)
\be
\tan\beta=\frac{v^s_{\perp}}{v^s_{||}}=\frac{\mu_{\perp}mg\cos\alpha}{\mu_{||}mg\sin\alpha}\simeq \frac1{2\tan\alpha}=\frac{\tan\alpha'}2
\ee 
where $m$ is the cylinder mass, $g$ the gravitationnal acceleration and $\alpha$ the angle between the symmetry axis and the horizontal ($\alpha'=\pi/2-\alpha$). The deviation angle with respect to the vertical is given by~\cite{guyon}
\be
\alpha'-\beta =\frac{\pi}2-\alpha-\beta \simeq \arctan\left(\frac{\tan\alpha'}{2+\tan^2\alpha'}\right)
\ee
We thus recover the result that for a a cylinder parallel ($\alpha=\pi/2$) or perpendicular ($\alpha=0$ ) to gravity, it falls vertically ($\beta=0$). The maximum fall angle is obtained for $\tan \alpha' \simeq \sqrt2$ and is around $\alpha'-\beta \simeq 20^{\circ}$. 

Now we assume that the object is a circular helix with its axis pointing along $\hat{\bz}$. 
We choose it arbitrary to be \textit{left-handed} (see Figure~\ref{helixBerg}b). Planes perpendicular to the centerline are no more symmetric planes and the coupling mobility $\bmu_{rt}$ is non-zero. The rotation of the helix at angular velocity $\bom=\omega \hat{\bz}$ (in Fig.~\ref{helixBerg}b we arbitraly choose $\omega>0$) can now induce a thrust $\mathbf{F}_T =\mu_{tr} \omega\hat{\bz}$. A simple local argument of the generation of the thrust is given by H. Berg~\cite{berg} and schematically shown in Fig.~\ref{helixBerg}b. We approximate the helix filament by a chain of small cylinders, which make an angle $\delta=2\pi R/\Lambda$ with the helix axis, where $R$ is the helix radius and $\Lambda$ the pitch. Because a cylinder has two different mobilites, Eq.~(\ref{SB}), the viscous drag is twice as big when it moves sideways than when it moves end-on. Hence the components of the drag normal to the helix axis contribute to the torque whereas components of the drag parallel to the axis are all of the same sign and contribute to a propulsive thrust $\mathbf{F}_T=F_T\hat{\bz}$ with $F_T>0$. Note that with our sign convention in Fig.~\ref{helixBerg}b, we find $\mu_{tr}>0$. This sign is related to the helix handedness: a right-handed helix would give $\mu_{tr}<0$. The velocity is $\bv=\bom \times \mathbf{R}$ and we find for the tangential and normal forces to the filament per unit length
\bea
F_{\perp}	&=& \frac{\omega R\cos\delta}{\tilde{\mu}_{\perp}} \\
F_{||}	&=& \frac{\omega R\sin\delta}{\tilde{\mu}_{||}}
\eea
where $\tilde{\mu}_{||}^{-1}$ and $\tilde{\mu}_{\perp}^{-1}$ are the friction cefficients per unit length. The thrust is then
\bea
\mathbf{F}_T &=& (F_{\perp}\sin\delta - F_{||}\cos\delta) \,\hat{\bz} \nonumber\\ &=& \frac12 \sin2\delta \,\left(\frac1{\tilde{\mu}_{\perp}}-\frac1{\tilde{\mu}_{||}} \right) R\, \bom
\eea
Hence the thrust is maximized for $\delta=\pi/4$. By approximating $\tilde{\mu}_{||} \simeq 2\tilde{\mu}_{\perp}$, we finally find the coupling mobility 
\be
\mu_{rt} \simeq \frac{4\tilde{\mu}_{\perp}}{R\sin2\delta}
\ee
Similarly, the rotation mobility associated with the resisting torque is given by 
\be
\mu_r \simeq -\frac{4\tilde{\mu}_{\perp}}{R^2(3+\cos2\delta)}
\ee

These approximate results allow a simple understanding of the propulsion mechanism. They are found using the so-called \textit{resistive force theory} developed by Gray-Hancock~\cite{gray} and Lighthill~\cite{lighthill,lighthill2,lighthill3}. The roughest approximation is the neglect of hydrodynamic interactions between different segments which is of course not justified since these interactions are long-ranged in $1/r$. Lighthill improves this approach by showing that these hydrodynamic interactions can be asymptotically calculated in the limit of slender body theory~\cite{cox,batchelor}, the small parameter being $a/q$ where $q\ll\Lambda$ is the unknown typical length of small cylinders and $a$ their thickness. Hydrodynamics are then taken into account by considering a helical line of Stokeslets for the far region ($r>q$) plus Doublets to fulfill the no-slip condition for the close region ($r<q$) where the helix is approximated by a straight filament. This allows a determination of $q=0.09\Lambda$ and local resistance coefficients per unit length $\tilde{\mu}_{||}^{-1}=2\pi\eta/\ln(2q/a)$ and $\tilde{\mu}_{\perp}^{-1}=8\pi\eta/[2\ln(2q/a)+1]$. Other works tackle the question of end effects and the influence of the cell body on the flow field, by considering Stokeslets, Doublets and a Rotlets which represents a point torque~\cite{ramia,Higdon}. Of course these studies suppose infinitely stiff helices. 

\section{Hydrodynamics of soft rods}
\label{soft}

In the following, we consider \textit{elastic} rods out of equilibrium under external forces. Contrary to rigid rods, hydrodynamic effects lead to a bending of elastics rods and therefore a reduction in symmetry~\cite{Elving,Mazur}. It results in an hydrodynamic translational-rotational coupling since now $\bmu_{rt}\ne 0$, which can be samll but coupled to large forces and torques can lead to very different qualitative behaviour compared to infinitely rigid rods. Indeed, sedimenting rods orient perpendicularly to the direction of motion due to a hydrodynamic torque. Likewise, the bending of rotating rods can induce propulsion in the same way as helices, contrary to rotating stiff cylinders for which the frictional force is always perpendicular to the rotation axis. Hence a bent rod can play the role of propeller.

\subsection{Orientation of elastic rods under external field}
\label{orient}

We consider an elastic rod submitted to an external force $F^\mathrm{ext}=qeE$ which can be the gravitational field in sedimentation experiments or an electric field ($E$) acting on a charged rod in birefringence experiments~\cite{xaver}.
As briefly described in the introduction, the orientation of stiff charged rods such as tobacco-mosaic viruses in electric fields is mostly determined by polarizability effects.
The anisotropic electric polarizability favors in general an orientation with the direction of the largest polarizability parallel to the external field $\bE$. 
The largest contribution to the polarizability comes from the easily deformable counterion cloud close to each charged rod, which is maximal along the rod axis~\cite{Yoshida,kikuchi,Netz,Netz1}. As a result, at not too low fields, charged rod-like particles orient parallel to $\bE$. However, when hydrodynamic effects are dominant and overwhelm this induced dipole mechanism, rod like particles can orient perpendicular to the field $\bE$. In the following, we focus on this hydrodynamic orientation. 

\begin{figure}[ht]
\begin{center}
\includegraphics[height=10cm]{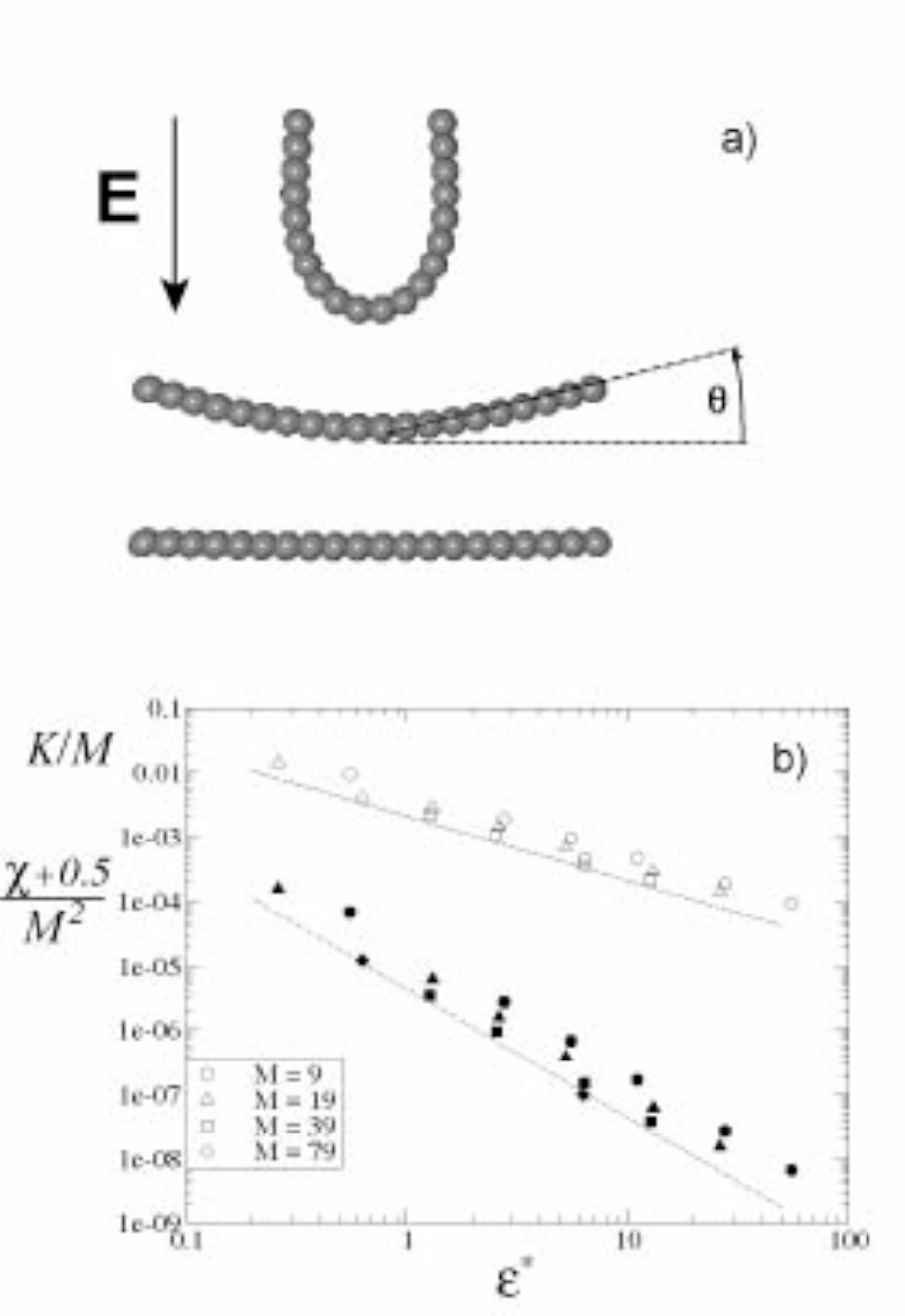} 
\caption{Stationary deformation and orientation of an isotropic elastic rod at \textit{zero temperature}. a)~Snapshots for $M=19$ and $\varepsilon^*$=0.03, 0.3 and 3 (from top to bottom) of a rod moving downwards. b)~Rescaled mean curvature $K/M$ (open symbols) and alignment parameter $(\chi+0.5)/M^2$ as a function of the rescaled rigidity $\varepsilon^*$ confirming scaling predictions Eq.~(\ref{scaling}) (broken lines with slopes -1 and -2).} 
\label{figX1}
\end{center}
\end{figure}

The mean chain bending is measured by $K=|\hat{\br}_{01}-\hat{\br}_{N\!-\!1\,N} | /2 = \sin\theta$ where $\theta$ is the bending angle of the terminal chain segment (see Fig.~\ref{figX1}a) and $\hat{\br}$ are the normalized bond vectors. The {\em overall} rod orientation can be measured by the {\em orientation parameter} $\psi$ which is defined as
\be
\label{eq_psi} 
\psi = \frac32 \left(\frac{\mathbf{R}_e\cdot \bE }{R_e E} \right)^2 -\frac12 = \frac{3\sin^2\alpha-1}{2}
\ee 
where ${\bf R}_e\equiv\br_{N-1}-\br_0$ is the end-to-end vector of the chain, and the orientation angle $\alpha$ is the same as in Fig.~\ref{helixBerg}a. The electric birefringence signal, on the other hand, is proportional to the {\em alignment parameter}~\cite{Elving}
\be 
\label{eq_birefringence} 
\chi = \frac1{M} \sum_{i=0}^{M-1} \left[ \frac32 \left( \frac{\br_{i i+1}\cdot \bE}{r_{i i+1} E}\right)^2  - \frac12 \right] 
\ee  
which is a measure of the average orientation of individual bonds and only equal to $\psi$ for straight rods. The hydrodynamic orientation mechanism, which of course only works at low Reynolds number, can be explained in simple terms as follows: 
the external force drives all monomers in the same way; the effective force, which results from the product of the hydrodynamic interaction tensor times direct forces, cf. Eq.(\ref{langevin}), is larger in the middle than at the two ends, because the middle receives hydrodynamic thrust from both sides. Since in stationary motion, all monomer velocities have to be the same, this imbalance in driving thrust is balanced by an {\em elastic deformation}, which in turn leads to an {\em orientation} of the rod.

On a scaling level, the elastic torque of a rod of length $L=2aM$ bent by an angle $\theta$, as shown in Fig.~\ref{figX1}a, is $N_\theta \simeq \varepsilon \theta/L$ and is balanced by an hydrodynamic torque, $N_h$, which arises from the above-mentioned inhomogeneity of hydrodynamic thrust. $N_h$ is thus proportional to the direct force per bead, the number of monomers $M$ and the rod length~: $N_h\simeq L^2qeE/a$. Equating $N_\theta \simeq N_h$ yields a stationary bending angle 
\be \label{scaling}
\theta \simeq \frac{L^3 qeE}{a\varepsilon} \simeq \frac{L}{a}\frac1{\varepsilon^*}
\ee
where $\varepsilon^*=\varepsilon/(qeEL^2)$ is the proper scaling variable in the continuum limit ($a\to0$). In Fig.~\ref{figX1}b, we show the numerically determined rescaled chain bending $K/M$ (open symbols) and the bond alignment $(\chi+1/2) / M^2$ (dark symbols) as a function of the inverse driving force $\varepsilon^*$ at {\em zero temperature} (the Langevin forces are switched off). Data are well described by the laws $\theta \simeq K = 1.1 \times 10^{-3} L/(a\varepsilon^*)$ and $ \theta ^2 \simeq \chi  +1/2= 1.1 \times 10^{-6} L^2 /(a\varepsilon^*)^2$ (broken lines), for not too strong bending, in agreement with our scaling predictions.

Rod bending reduces the symmetry and hydrodynamically couples translational and rotational degrees of freedom; due to a shift between the center of mass and hydrodynamic stress the bent rod is oriented perpendicularly to the direction of motion with the opening pointing backwards \cite{happel}. For finite temperature the orientation is subject to thermal fluctuations and thus not complete. If we consider an assembly of elastic rods in solution as in sedimentation or birefringence experiments, we can deduce the average bending and orientation using Boltzmann statistics. Within linear response, the orienting torque $N_\alpha$ is proportional to the orientation angle $\alpha$, the bending angle $\theta$ and the driving torque $N_h$: $N_\alpha \sim \alpha \theta N_h$. 

\begin{figure}[ht]
\begin{center}
\includegraphics[height=10cm]{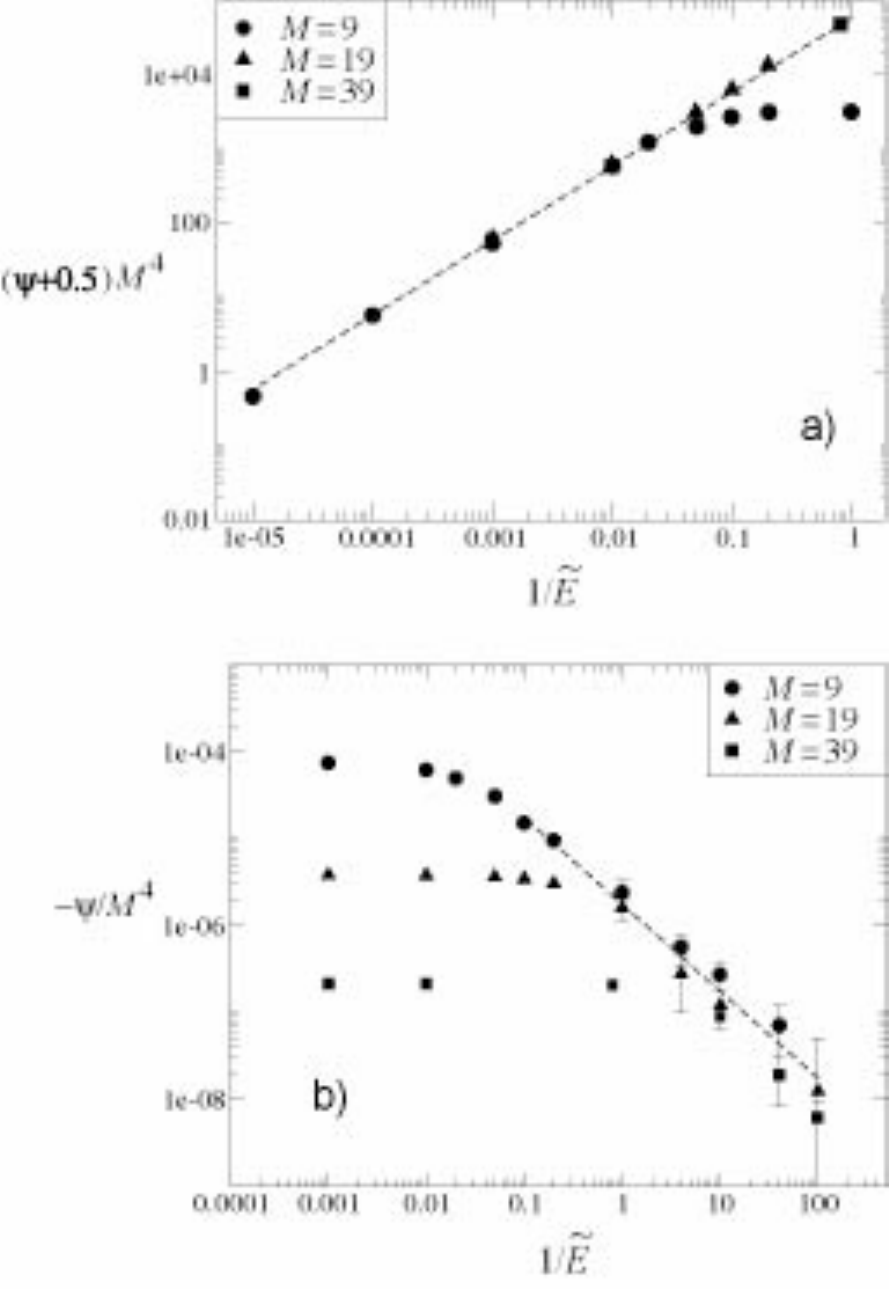} 
\caption{a)~Rescaled orientational parameter $(\psi+0.5) M^4$ as a function of the  rescaled temperature $k_BT/(qeEa) = 1/\tilde{E}$ for fixed rigidity $\varepsilon^*L/a=100$ in agreement with the low-temperature scaling prediction Eq.(\ref{lowtpsi}) (broken line). b)~Same data plotted as $-\psi/M^4$ compared with the high-temperature scaling prediction Eq.(\ref{hightpsi}).} 
\label{figX2}
\end{center}
\end{figure}

For {\em low temperatures}, the average orientational energy equals the thermal energy, $\alpha N_\alpha \sim k_BT $, leading to a mean square orientational fluctuation
\be \label{lowtpsi}
\frac23 \left(\psi +\frac12\right) \simeq \langle \alpha^2 \rangle = 6.1\times 10^3 \left(\frac{a}{L}\right)^3 \frac{\varepsilon^*}{\tilde{E}}
\ee
where $\tilde{E}$ is defined in Eq.~(\ref{tildeE}). It is checked for a fixed rescaled chain stiffness $\varepsilon^*L/a =  \ell_\mathrm{p}/(L\tilde{E})$ in Fig.~\ref{figX2}a with a numerically determined prefactor. The orientational fluctuations at linear response are governed by a quadratic form proportional to a coupling constant $J$ with expectation value
\begin{equation} \label{eqeq}
\langle \sin^2 \alpha \rangle =  -\frac{\partial}{\partial (J/2)} \ln
\int_{-\pi/2}^{\pi /2} d \alpha \cos \alpha \, e^{- \frac{J}2 \sin^2\!\alpha}
\end{equation}
For low temperatures (high $J$-values) one obtains $\langle \sin^2 \alpha \rangle \simeq \langle  \alpha^2 \rangle = 1/J$ which together with Eq.~(\ref{lowtpsi}) fixes the coupling constant $J$. 

The {\em high-temperature} regime is experimentally more relevant and corresponds to an almost isotropic rod orientation distribution. In this limit, one obtains from Eq.~(\ref{eqeq}) $\langle \sin^2 \alpha \rangle \simeq 1/3 - 2 J /45 +{\cal O}(J^2)$, which together with the definition of $\psi$, Eq.~(\ref{eq_psi}) yields the final result
\begin{equation}  \label{hightpsi}
\psi \simeq - \frac{J}{15} = -1.1 \times 10^{-5} \tilde{E}^2 \frac{L}{\ell_\mathrm{p}} \left(\frac{L}{a}\right)^4
\end{equation}
Equation~(\ref{hightpsi}) including the numerical prefactor is confirmed in Fig.~\ref{figX2}b (broken line). As might be expected for the stationary behavior, the result is independent of the solvent viscosity. The dependence on the rod length $\psi\sim L^5$ is stronger than for a mechanism based on electric counterion polarizability, where one typically assumes an $L$-dependence like $\psi\sim L^3 E^2$ or even smaller~\cite{Yoshida}. This means that hydrodynamic orientation is dominant for long rods, in agreement with experimental observations showing increased anomalies for long polymers~\cite{Opper}. Finally using Eq.~(\ref{scaling}) and (\ref{hightpsi}), the ratio between chain orientation and chain bending becomes $|\psi| / \theta^2  \simeq \ell_\mathrm{p} /L$, which suggest that for stiff polymers $\ell_\mathrm{p} > L$ one can observe weakly bent but still strongly oriented chains.

Transient birefringence and dichroism experiments~\cite{Fredericq} deal with the {\em time dependence} of the orientational response $\chi(t)$ and $\psi(t)$ after the electric driving field, $F^{\mathrm{ext}}=qeE$ is suddenly turned off (or on). The orientational diffusion time of a rod, $\tilde{\tau}_D  = \tau_D  q eE \mu_0 /a= 218$, is given by the perturbative expression for the rotational diffusion constant~\cite{Tirado} in powers of $a/L$. At the simplest level, the decay of the orientation angle $\alpha$ in an applied electric field follows a drift orientational diffusion and is governed by the equation 
\be
\frac{\mathrm{d}\alpha}{\mathrm{d}t} = - \frac{D}{k_BT} N_\alpha
\ee
where $N_\alpha$ is the orientational torque. The hydrodynamic orientation time is therefore
\be
\tau_{\mathrm{hd}} = \frac{k_BT}{D\theta N_h} \simeq \frac{a\varepsilon}{\mu_0 (qeE)^2  L^2}
\ee
hence within logarithmic corrections $\tau_D / \tau_{\mathrm{hd}}  = 1.1 \times 10^{-5}  L^5 (qeE)^2 /(a^2 \varepsilon k_BT)$. The strong length dependence makes the orientational process much faster for long rods.
For charged rod-like objects, the hydrodynamic orientation mechanism competes with the counterion polarization mechanism which favors parallel orientation. The orienting polarization torque is $N_\alpha^P  \sim L^3 \varepsilon_0 \alpha E^2$ where $\varepsilon_0$ is the dielectric constant (neglecting the dispute about the $L$ dependence of the polarizability~\cite{Yoshida}). The characteristic orientation time due to electric polarization, $\tau_P$, follows from a similar equation as above, $\tau_{P} \sim (k_BT/D) (\alpha/N_\alpha^P)\sim 1/(\varepsilon_0 \mu_0 a E^2)$, and is independent of the length. Hence, for long polymers one has $\tau_D > \tau_{P} > \tau_{\mathrm{hd}}$, and the hydrodynamically induced anomalous birefringence is the fastest process. 

We illustrate the proposed mechanism with fd-viruses which have a length of $L \simeq 880$~nm and a diameter of $2a \simeq 9$~nm. The total net charge is roughly $ 500 \,\,e^-$ so that the valency per length $2a$ is about $q = 5$~\cite{kramer}. For a  typical electric field $E = 10^5$~V/m the rescaled field strength is $\tilde{E} \simeq 0.1$. Equation~(\ref{hightpsi}) yields an  orientation of the order of unity when $L/\ell_\mathrm{p} > 10^7 (a/L)^4 \sim 0.1$, a realistic number for a virus material. In the case of sedimentation, the reduced driving field acting on the monomers is $\tilde{E} \simeq 4 \pi a^4 \rho g G/(3 k_BT)$ where $G=9.81$m/s$^2$ is the gravitational acceleration, $g$ is the g-factor of a centrifuge, and $\rho \simeq 10^3$~kg/m$^3$ is the density difference between the sedimenting particle and the solvent. For a particle radius $a \simeq 10^{-8}$~m, we obtain $\tilde{E} \simeq 10^{-7}g$. To obtain the same effect as in the above charged-rod example, one would need a $g$-factor of $g \simeq 10^6$ which is large but reachable in an ultracentrifuge. For larger colloidal rod-like particles,
sedimentation will lead to hydrodynamic orientation even without centrifuging: large particles are predicted to sediment perpendicularly to the direction of motion~\cite{Buiten}.

All our results are valid only for  low Reynolds numbers: for an object of length $L$ moving at velocity $u$, the Reynolds number is $Re\sim L \rho u /\eta$. For a cylinder the velocity scales as $u \sim F/\eta L$ where the total force is $F \sim q e E L $. The condition $Re < 1$ leads to $(L/a) (a qe E/ k_BT) \sim N \tilde{E}  < \eta^2 /(\rho k_BT)  \approx 10^{11}$ (where we used the density and viscosity of water) which is verified in our simulations. Hence the hydrodynamic orientation discussed here is compatible with low Reynolds number hydrodynamics.

\subsection{Propulsion with a rotating elastic rod}
\label{prop}

We have seen in the previous section that the coupling between elasticity and hydrodynamics leads to surprising effects: a straight elastic rod that is sedimenting in a quiescent fluid and thus subject to a homogeneous external force distribution becomes bent and oriented  perpendicular to the force. 
Here we investigate a different scenario: if one rotates an elastic rod by application of a torque at one of its ends, it will bend due to frictional forces. In the absence of bending, i.e for an infinitely stiff rod, a simple symmetry argument shows that the net thrust when averaged over a full turn vanishes.
Clearly, a rod fixed at one end which rotates making a tilt angle with the rotation axis in the direction $\hat{\bz}$, will induce no thrust along $\hat{\bz}$. The cylinder moves sideways and the frictional force, given by $\mathbf{F}=\mu_{\perp} \bom\times\br$, is always perpendicular to $\bom=\omega\hat{\bz} $. However, if the rotating rod is elastic, the friction forces will bend the rod
  and the question of propulsion thrust can be raised again. 
A few theoretical works have focused on the coupling between hydrodynamics in a viscous medium and elasticity of soft polymers~\cite{wiggins,camalet,goldstein}. It was shown that finite stiffness of 
beating straight filaments breaks the time-reversal symmetry and enables propulsion.
Self-propulsion using alternative mechanisms such as surface
distortions~\cite{Stone} or chemical reactions~\cite{Ajdari} has been actively
proposed for designing macromolecular machines working in a viscous media.
In this section, we present simulation results which indicate propulsion for rotating straight filaments~\cite{manoel}. 

\begin{figure}[ht]
\begin{center}
\includegraphics[height=4.5cm]{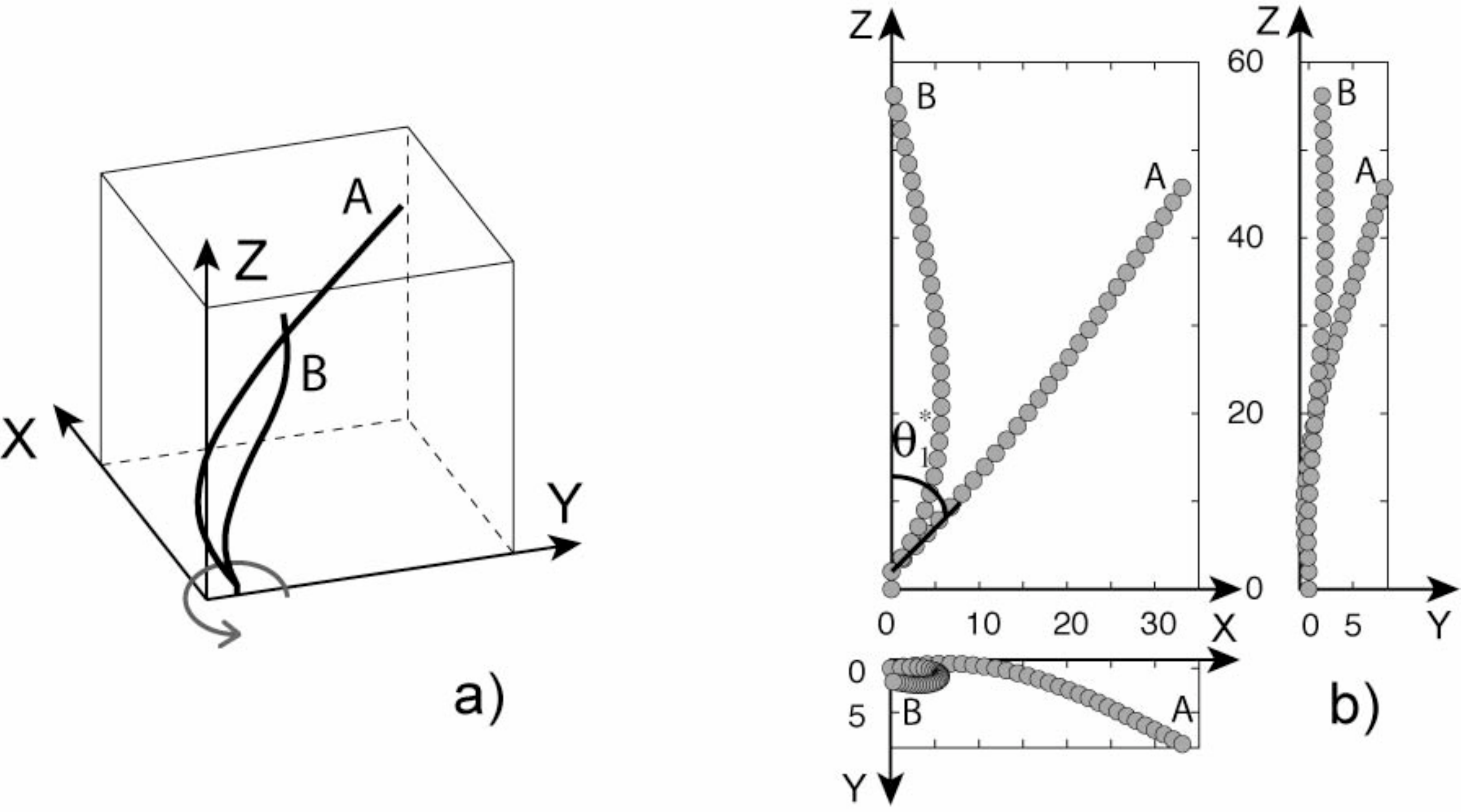} 
\caption{Sketch in 3D (a) and projected views (b) of the rotating filament before (A, $\tN=600$) and after (B, $\tN=800$) the bifurcation ($M=30$ beads, $\ell_{\mathrm{p}}/L=10^3$).}
\label{sketch}
\end{center}
\end{figure}

To mimic a rotary motor, we apply an external force on monomer 2 of a linear array
of monomers, which is related to the applied torque $\mathbf{N}=N\hat{\bz}$ by $\mathbf{F}^{\mathrm{ext}}=\mathbf{N}\times\mathbf{r}_{12}/r_{12}^2$. The geometry and sign convention are shown in Fig.~\ref{sketch}. Two different force ensembles are investigated: the \textit{stalled} case, where the first two monomers forming the polymer base are held fixed in space by applying virtual forces which exactly cancel all other elastic and hydrodynamic forces. In the \textit{moving} case, we let the polymer move along the $\hat{\bz}$ direction and we apply virtual forces only laterally such that the base moves along a vertical straight track. A finite tilt angle at the base is imposed by a spontaneous curvature term in the elastic energy $(\varepsilon/2a)[1-\cos(\theta_1-\theta^*_1)]$; here we show data for spontaneous curvature $\theta^*_1=45^{\circ}$. 

When a torque is applied to the filament base, it rotates and after a few turns 
exhibits a stationary shape. Due to friction the filament bends into a curved structure. 
It should be noticed that, contrary to Section~\ref{orient}, the applied force is not homogeneous along the polymer. However the angular velocity of the whole polymer is the same for every monomer. Hence, the velocity $\bv=\bom\times\br$, and therefore the frictional force, is larger at the end of the polymer, causing a stronger bending opposite to the direction of rotation. The bending due to hydrodynamic interactions (which are larger in the middle of the rod than at the ends), similar to effects studied in Section~\ref{orient}, is thus a secondary and negligible effect.

\begin{figure}[ht]
\begin{center}
\includegraphics[height=15cm]{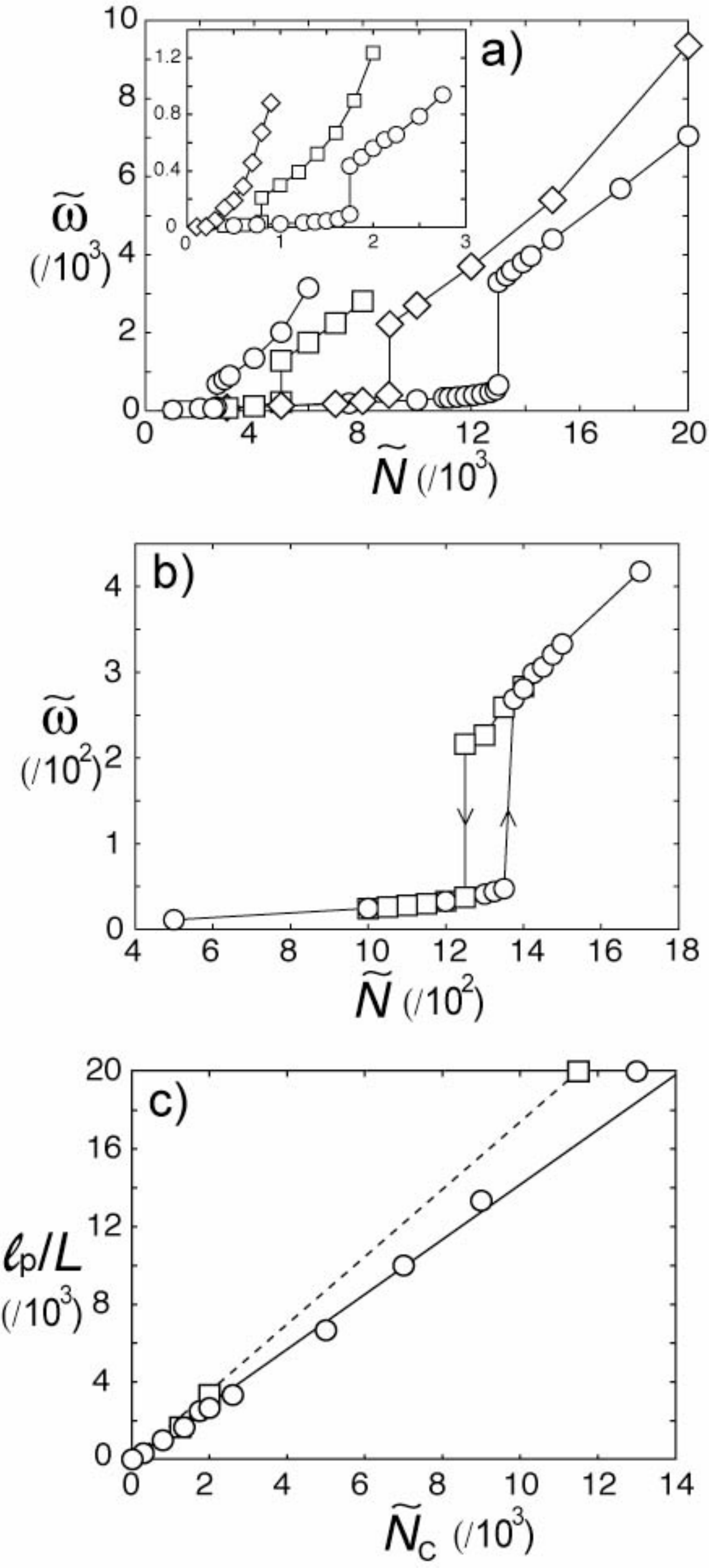} 
\caption{a) Angular velocity as function of external torque, $\tilde{N}$, for various $\ell_{\mathrm{p}}/L$: 3333 ($\circ$), 6667 ($\square$), $1.33\times10^4$ ($\diamond$) and $2\times10^4$($\circ$) [inset: 333 ($\diamond$), 1000 ($\square$) and 2500 ($\circ$)].~b) Hysteresis cycle for $\ell_{\mathrm{p}}/L=1667$ for increasing  ($\circ$) and decreasing torque ($\square$) at constant rate $\delta \tilde{N}/(\delta t/\Delta)=1.25\times10^{-6}$.~c) Persistence length vs. critical torque for increasing ($\circ$) and decreasing  torque ($\square$) both obeying a linear law according to Eq.~(\ref{scaling prediction}) (All data for stalled case, $V_z=0$).}
\label{omega}
\end{center}
\end{figure}

The stationary angular velocity $\tilde{\omega}=\omega \Delta/\tilde{\mu_0}$ (in degrees), as a function of the applied torque $\tilde{N}=N/k_BT$, is plotted in Fig.~\ref{omega}a. It shows a non-linear increase, which is associated to the finite bending rigidity. Indeed, for an infinitely stiff polymer, this relation would be strictly linear due to Stokes flow properties. For a critical torque, $\tilde{N}_{\mathrm{c}}$, a shape bifurcation occurs and the angular frequency jumps dramatically. Figure~\ref{sketch} shows the stationary shape of the polymer before (A) and after the bifurcation (B). The high-torque state is characterized by a smaller distance $r_\perp$ from the rotation axis. This explains the observed jump in angular velocity since the applied torque, \textit{i.e.} the frictionnal forces, remain constant at the bifurcation. Since the velocity of a bead $i$ is $\bv_i=\bom\times\br_i$, we find $\omega=v_i/r_{\perp i}$ which jumps dramatically since $r_{\perp i}$ decreases.
Pronounced hysteresis, which becomes slightly weaker with decreasing torque sweep rate, is observed when $\tilde{N}$ is varied across the critical region, as illustrated in Fig.~\ref{omega}b. This hysteresis cycle indicates that this shape bifuraction is subcritical.

What is the control parameter of this bifurcation? 
 Figure~\ref{omega}c shows the dependence of the critical torque on the persistence length of the rod. Since the torque sweep rate is finite, increasing ($\circ$) and decreasing torque ($\square$) give rise to slightly different bifurcation values. For very small persistence lengths, conformational fluctuations suppress the transition. The shape bifurcation can be understood by simply balancing elastic and driving torques: the bending torque due to the filament deformation, projected along the vertical axis, reads $\sim \varepsilon\sin\theta_1/R$ where $R$ is the bending radius of the filament, which has to be balanced by the external torque $N$. The onset of the transition is fixed when the bending radius reaches the length of the filament, $R \sim L$. This yields a critical torque 
\begin{equation}
\tilde{N}_{\mathrm{c}}\simeq\frac{\ell_{\mathrm{p}}}{L}\sin\theta_1
\label{scaling prediction}
\end{equation}
which is shown in Fig.~\ref{omega}c as a solid line and agrees very well with the numerical data. It transpires that the bifurcation is purely \textit{elastic} in origin, since neither hydrodynamical parameters nor temperature appear in Eq.~(\ref{scaling prediction}). Using the rotational mobility $\mu_{\mathrm{rot}} \sim 1/(\eta L^3)$, the lower critical angular frequency reads 
$\omega_{\mathrm{c}} \sim \mu_{\mathrm{rot}} N_c  \sim \varepsilon \sin \theta_1 /(\eta L^4)$. This threshold turns out to be much lower than for  the continuous twirling-whirling transition of a rotating rod with torsional elasticity, $\omega_{\mathrm{c}}^{\mathrm{TW}} \sim \varepsilon/(\eta a^2  L^2)$, which was obtained using linear analysis~\cite{goldstein}. Moreover, the twirling-whirling transition is related to a symmetry breaking, named as supercritical bifurcation.

\begin{figure}[ht]
\begin{center}
\includegraphics[height=4.5cm]{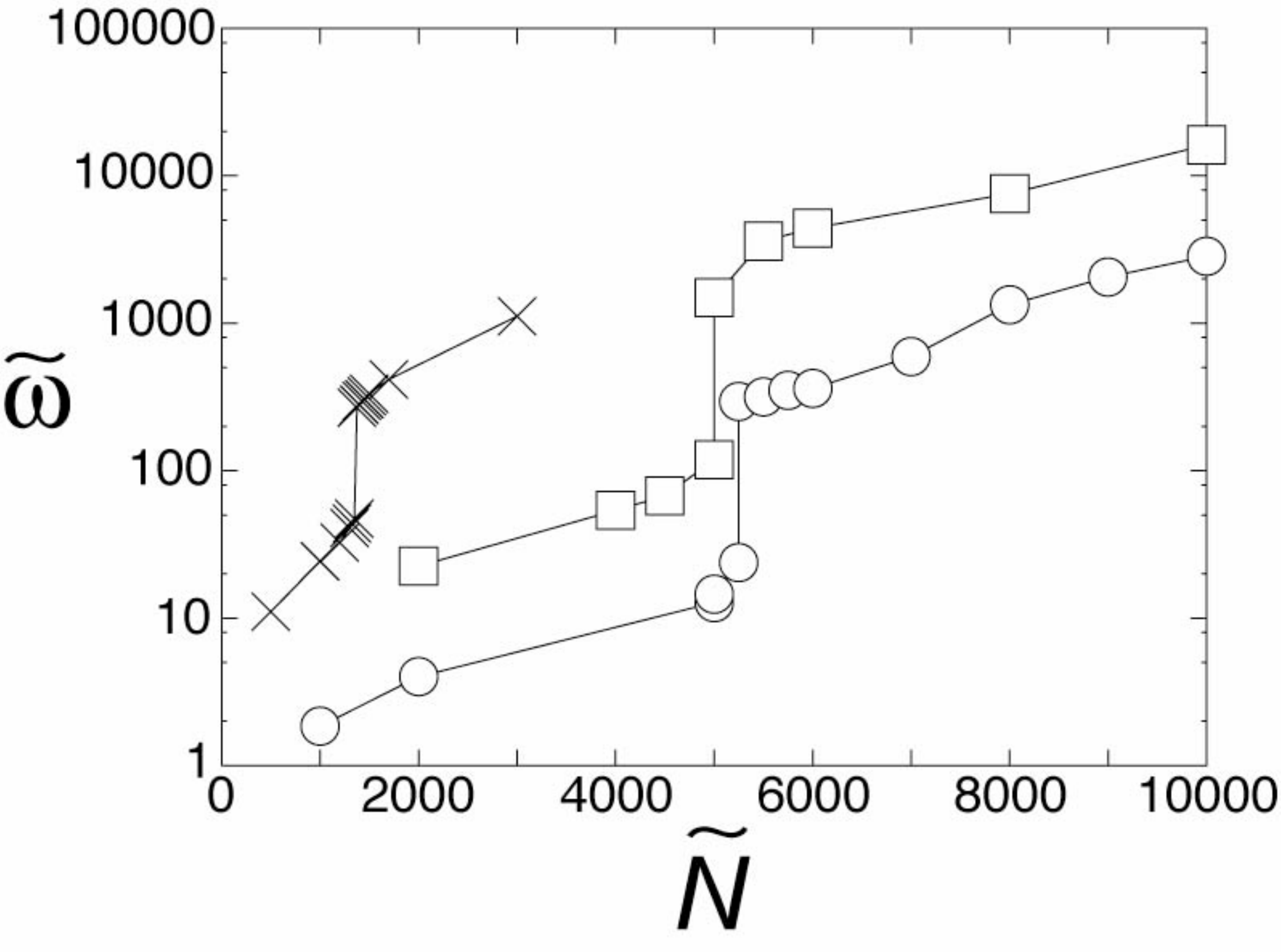} 
\caption{Angular velocity vs. applied torque for $\ell_{\mathrm{p}}/L=1667$ using 
full hydrodynamic interactions ($\times$), slender-body ($\square$) and free-draining approximations ($\circ$). The transition occurs in all cases but hydrodynamics decrease significantly the critical torque (Stalled case, $V_z=0$).}
\label{3cases}
\end{center}
\end{figure}

At this point, one may wonder what the role of hydrodynamic interactions is in this transition. This role is twofold: first, it induces a propulsive thrust, since we have seen in Section~\ref{rigid}, that hydrodynamic interactions are essential in the production of thrust by a rotating helix. Obviously, our bent filament for large torques does not have the geometry of a pure helix but it wraps around the rotation axis. The second aspect is more subtle. To understand the role of hydrodynamic interactions for the shape bifurcation, we implemented two additional approximations: 
\begin{itemize}
	\item the free draining approximation (FDA), where hydrodynamics are ignored altogether. The mobility matrix is thus isotropic: $\mu_{ij} = \mu_0\; \mathbf{1} \;  \delta_{ij}$;
	\item the resistive force theory within the slender body approximation (SBA) which is defined in Section~\ref{rigid}~\cite{lighthill}. The mobility matrix is given by Eq.~(\ref{SB}). The choice of the values of the two local mobilities in the simulations is of course difficult and heuristic, since at our coarse-grained scale the monomers are spheres which sets $\mu_{\perp}=\mu_{||}=\mu_0$. Values for long cylinders or long ellipsoids are given in literature and we arbitrarily chose $\mu_{\perp}=3\mu_0 (2\ln M +1)/4$ and $\mu_{||}=3\mu_0 (2\ln M -1)/2$. Note that these values should in principle be adjusted to the shape of the polymer.
\end{itemize}
Figure~\ref{3cases} shows the angular velocity versus the applied torque, $\tilde{N}$, for the three approximations. In accord with our elastic scaling model, the transition occurs for all approximations even in the absence of hydrodynamics (FDA: $\circ$). However, the critical torque is overestimated for both FD and SB approximations by roughly a factor of 5. 

\begin{figure}[ht]
\begin{center}
\includegraphics[height=15cm]{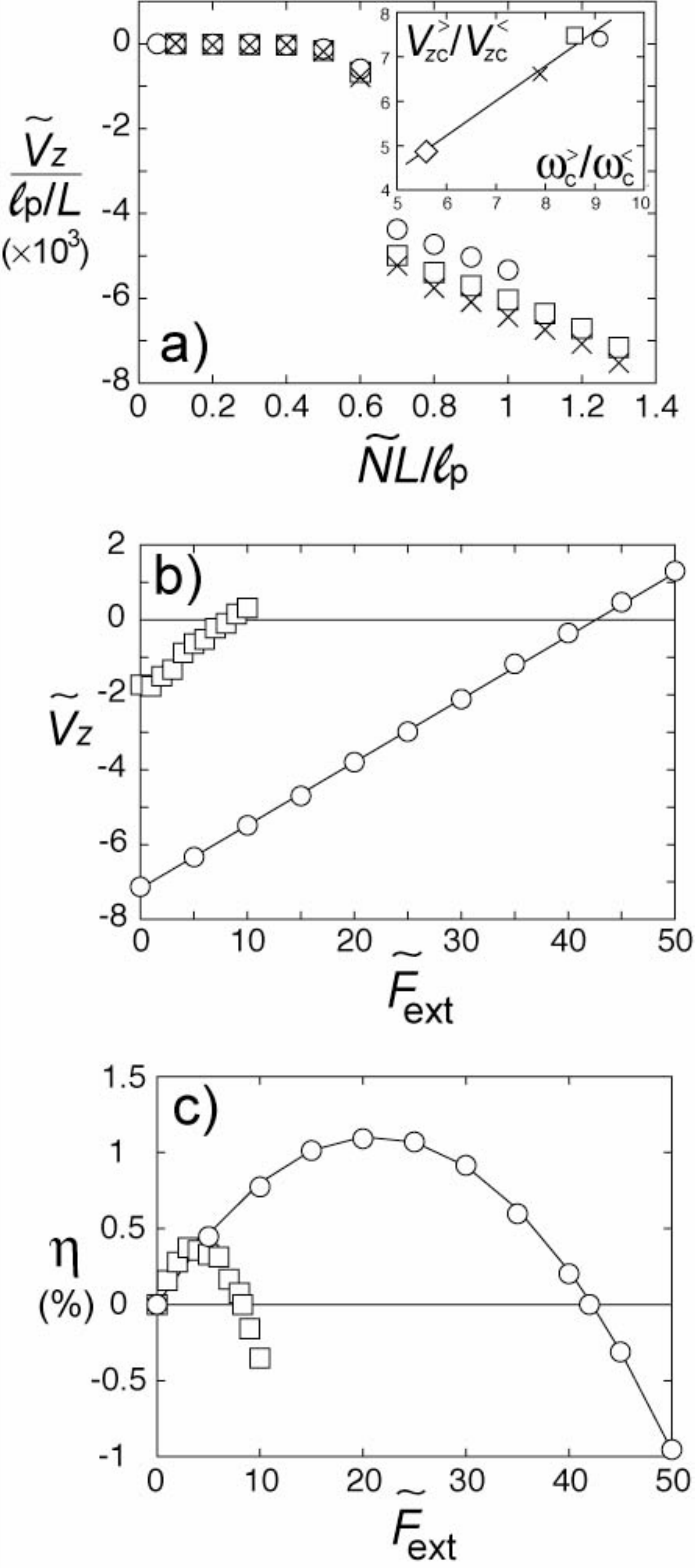} 
\caption{a) Propulsion velocity parallel to the rotation axis vs. $\tilde{N}L/\ell_{\mathrm{p}}$ in the $z$-\textit{moving} case for $\ell_{\mathrm{p}}/L=2500$ ($\times$), 5000 ($\square$) and $2\times10^4$ ($\circ$). The inset shows the variations of the ratio of velocities after and before the transtion vs. the ratio of angular velocities ($\diamond$, $\ell_{\mathrm{p}}/L=1667$). b) $\tilde{V}_z$ vs. the external force $\tilde{F}_{\mathrm{ext}}$ applied on monomer 2 ($\ell_{\mathrm{p}}/L=1667$) before the bifurcation for $\tilde{N}=0.78 \tilde{N}_{\mathrm{c}}$ ($\square$) and after $\tilde{N}=0.84 \tilde{N}_{\mathrm{c}}$ ($\circ$). c) Efficiency, $\eta$, vs. $\tilde{F}_{\mathrm{ext}}$ (same parameters as b). The solid line is a  parabolic fit.}
\label{propulsion}
\end{center}
\end{figure}

Is it possible to define the mobilities of the whole object? The relation between angular velocity and torque is no more linear and the meaning of the matrix in Eq.~(\ref{mobilities}) changes. We define four mobilities of the whole object following Eq.~(\ref{mobilities}). The dimensionality of the mobility matrix is now $2\times 2$ since we consider only the $z$-component of torques, forces and velocity of the whole rod:  $\bv=V_z\hat{\bz}$ and $\mathbf{F}=F_{\mathrm{ext}}\hat{\bz}$ the corresponding external force applied at the propeller base. Nevertheless, we shall keep in mind that the matrix notation is just used for an easier comparison to rigid propellers, because these four mobilities ($\mu_\mathrm{rr}$, $\mu_\mathrm{rt}$, $\mu_\mathrm{tr}$ and $\mu_\mathrm{tt}$) depend \textit{a priori} on $N$ and $F^\mathrm{ext}$. Moreover, for a flexible rod the symmetry of the mobility tensor is not satisfied. The propulsion velocity along the rotation axis is plotted in Fig.~\ref{propulsion}a as a function of  the control parameter $\tilde{N}L/\ell_{\mathrm{p}}$ for different persistence lengths for the $z$-\textit{moving} case, i.e with $F_{\mathrm{ext}}=0$. At the transition ($\tilde{N}L/\ell_{\mathrm{p}}\simeq 0.7$), a jump in the propulsion velocity is observed and $\tilde{V}_z$ is almost linear for $\tilde{N}>\tilde{N}_{\mathrm{c}}$, revealing that the shape remains almost fixed in this range of torque values. Inset of Fig.~\ref{propulsion}a shows the ratio of $V_z$ before and after the transition against the same ratio for angular velocities. The variation is roughly linear, meaning that the jump in propulsion is mostly due to the increase in angular velocity. A refined analysis would be necessary to detect if the rod takes a specific shape which favors the forward velocity. For instance, in the case of a perfect helix, we have seen that the thrust is maximised for a pitch angle equal to $\alpha=45^o$. We have checked that all the previous results for $\omega(N)$ are very similar in the $z$-\textit{moving} case. The only difference appears at the bifurcation, the critical torque being a little smaller without any external load force.

Up to now, our model neglected the presence of a base for the rotor.
In reality, the filament is attached at its end to a body, the head of the whole propeller. We assume that the relative rotation of both components is free. The net force and torque on the complete propeller (body + filament) swimming at constant velocity are zero~\cite{purcell} since the motion is overdamped preventing 
any acceleration. Hence, for a spherical body of radius $R$ and angular velocity $\Omega$, the loading force is proportional to the body velocity whereas the loading torque is proportionnal to $\Omega$, which reads
\bea
F_{\mathrm{ext}} &=& -6\pi\eta R\,V_z \label{Lmatching}\\
N &=& -8\pi\eta R^3\,\Omega \label{Lmatching2}
\eea
To describe completely the whole propeller with a perfect load matching, it is thus necessary to find the characteristic response of the filament to an external load. To test numerically our propeller under load, we applied an external force, $F_{\mathrm{ext}}$, which we define to be positive when it pushes against its natural swimming direction. Figure~\ref{propulsion}b shows the variation of $\tilde{V}_z$ with $\tilde{F}_{\mathrm{ext}}$ for $\ell_{\mathrm{p}}/L=1667$ at two different torques, just below and above the bifurcation. The laws are almost linear in both cases, meaning that $\mu_{\mathrm{tt}}$ is almost independent of $F_{\mathrm{ext}}$. Hence for given torque $N$ and body radius $R$, load matching fixes $F_{\mathrm{ext}}$ and thus $V_z$ by taking the intersection between the above law and Eq.~(\ref{Lmatching}). The angular velocity of the body is then found using Eq.~(\ref{Lmatching2}). The \textit{stalled} case corresponds to an infinite spherical body ($R\to\infty$) whereas the free $z$-\textit{moving} case imposes $R=0$, \textit{i.e.} no body. Of course, our study is in principle not valid for $R\gg 1$ because the presence of a hard wall will perturb the flow at the filament base. The efficiency of the power converter can be defined as the ratio of the propulsive power output and the rotary power input,
\begin{equation}
\eta=-\frac{F_{\mathrm{ext}} V_z}{N\omega}.
\label{efficiency}
\end{equation}
By inserting Eq.~(\ref{mobilities}) in Eq.~(\ref{efficiency}), we obtain 
\be
\eta(F_{\mathrm{ext}})=-\frac{\mu_{\mathrm{tr}}NF_{\mathrm{ext}}+\mu_{\mathrm{tt}} F_{\mathrm{ext}}^2}{\mu_{\mathrm{rr}}N^2+\mu_{\mathrm{rt}} N F_{\mathrm{ext}}}
\ee
We have checked that $\mu_{\mathrm{rt}}$ is negligibly small. Hence, the efficiency becomes parabolic as a function of the external force as shown in Fig.~\ref{propulsion}c. The highest efficiency is obtained for $F_{\mathrm{ext}}=-\mu_{\mathrm{tr}}N/(2\mu_{\mathrm{tt}})=F_{\mathrm{stall}}/2$ and is only of the order of 1\% after the transition and 3 times smaller before. With this load, we find a body size $R\simeq 5\,a$ and an unique $|\tilde{\Omega}|=\frac34 (a/R)^3 \tilde{N}\simeq8\ll\tilde{\omega}\simeq250$ for $\tilde{N}=0.84 \tilde{N}_{\mathrm{c}}$ after the transition.

In conclusion, we show that a simple straight elastic filament can induce propulsion when rotated at one end. The interplay between elastic deformations and hydrodynamic interactions results in a substantial directed thrust due to a subcritical dynamic bifurcation for a control parameter $\tilde{N}/\tilde{N}_{\mathrm{c}}$ larger than 1. The mobility $\mu_{\mathrm{tr}}$ changes its sign when the torque is reversed, which implies a forward thrust whatever the sense of rotation. 

This work provides a clue for the synthetic manufacture of biomimetic micro-propeller, by using simple semi-flexible polymers instead of rigid helices. This bifurcation could be experimentally investigated using a macroscopic scale model similar to the one developed by Powers group~\cite{kim}. With their experimental setup, the bifurcation would occur for torques on the order of 0.05~N.m and angular velocities larger than 0.01~Hz, which are accessible. Another type of experiments has been developed by Bibette's group, where they make microscopic artificial swimmers with chains of connected ferrofluid droplets swimming by applying a transverse oscillatory magnetic field~\cite{dreyfus}. The resulting forward thrust is due to a beating of this "elastic" chain. Even if these experimental results are associated to transverse beating, one can imagine quite soon a rotating chain experiment.

\subsection{Grafted elastic polymers in shear flow}
\label{shear}

In previous theoretical studies, the effects of shear flows on surface-anchored polymers 
have been primarily studied analytically and numerically for  dense brushes  and flexible 
polymers~\cite{Harden,Binder,Muller}. In this section, we study the shear  flow response of a layer of semiflexible polymers grafted  to a 
planar surface (a so-called polymer brush) as a function of polymer stiffness, grafting density
and shear rate~\cite{woon}. A finite orientational stiffness is introduced at
the grafting  points, 
tending to vertically orient the  grafted chains at the basal region.
The time evolution of the system is described by the Langevin equation~(\ref{langevin})
where a linear external shear flow
${ \bf v}({\bf r})=\dot{\gamma} z \mathbf{\hat{x}}$ along  the $x$ direction
with a shear rate $\dot{\gamma}$ is included.
We explicitly incorporate full hydrodynamic interactions between monomers
in the presence of a no-slip surface through a non-diagonal mobility tensor ${\bf \mu}_{ij}$ on the Rotne-Prager level~\cite{russel} following Section~\ref{surf}.
As before, we rescale  all lengths and  energies  and express
the rescaled shear rate by $\dot{\tilde{\gamma}}=a^3 \dot{\gamma} \eta /k_BT$.
In order to mimic a semi-infinite surface covered with
anchored chains at a finite density, 
we consider 4 polymers with monomer number $N=10$ 
grafted regularly within a  square
unit cell of lateral size $D$ and treat hydrodynamic
interactions \textit{via} lateral periodic boundary conditions
imposed on the flow field.
The grafting density is defined as $\rho = 4 a^2/D^2$.

\begin{figure}[ht]
\begin{center}
\includegraphics[width=8.5cm]{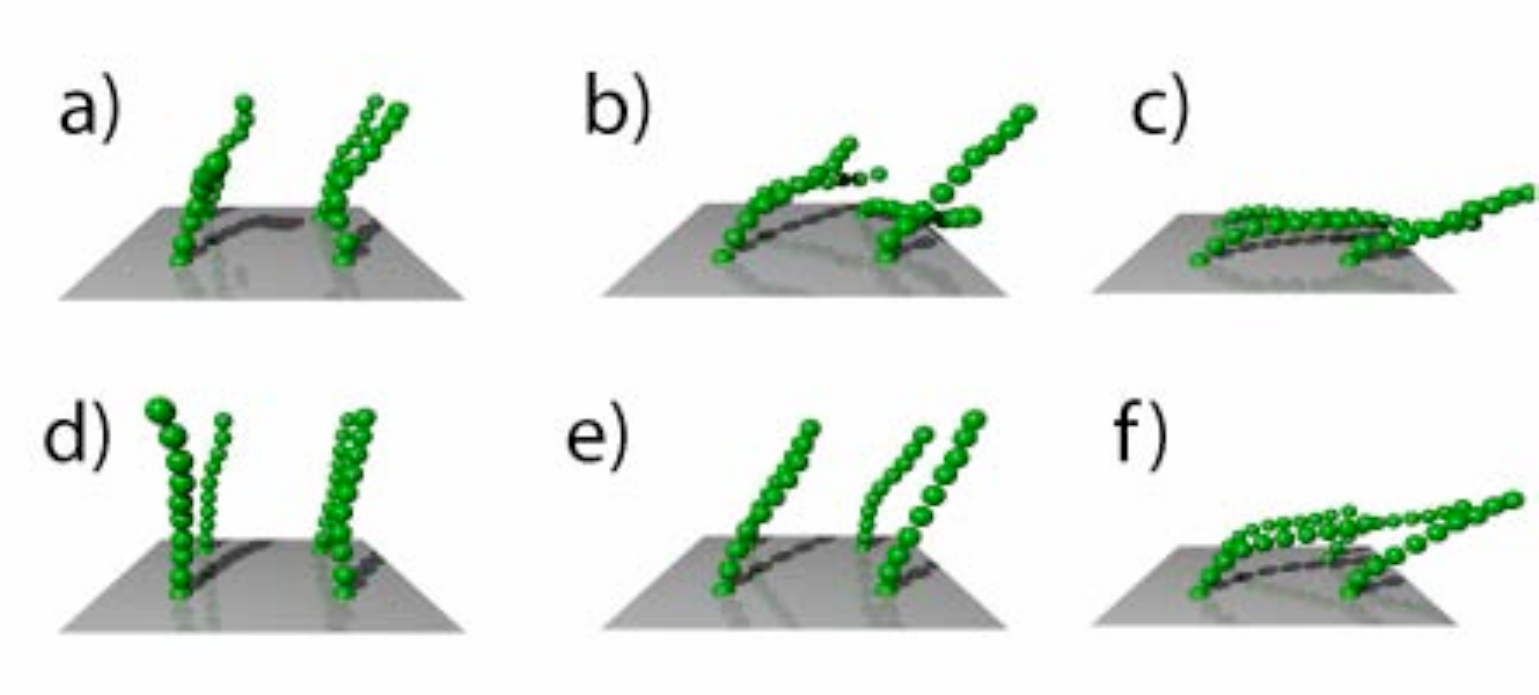}
\caption{Snapshots of polymer configurations for grafting density $\rho=0.005$ and semiflexible polymers (top row, persistence length $\ell_p/L =1$)and rather stiff polymers (bottom row, persistence length $\ell_p / L =5$). To the left no shear flow is acting, in the middle column the shear rate is $\dot{\tilde{\gamma}} =0.003$ (corresponding to a shear rate in unrescaled units of $\dot{\gamma} \approx 10^7$ s$^{-1}$ for sphere radius $a=1$nm and water as a solvent), and  to the right the shear rate is $\dot{\tilde{\gamma}} =0.03$ (corresponding to a shear rate in unrescaled units of $\dot{\gamma} \approx 10^8$ s$^{-1}$ for $a=1$ nm).}
\label{fig9}
\end{center}
\end{figure}

The effects of a shear flow is demonstrated in typical simulation snapshots shown 
in Fig.~\ref{fig9}. The upper row is for semiflexible polymers, while the bottom row
shows results for rather stiff polymers. As one goes from the left to the right, the
applied shear rate is increasing steadily, and one observes a clear change of the
polymer conformations.  
Upon shearing the brush, the polymers tend to be bent
to reduce hydrodynamic friction, and
in turn exert a resistive response to the flow due to a finite
bending stiffness.
Therefore, the resulting solvent flow velocity is expected to be reduced 
compared to the case of a bare
surface in the absence of grafted chains.
As the shear rate increases, the bending of the polymers becomes stronger
and they move closer to the surface.
As a result, the flow can approach the no-slip surface more closely.

\begin{figure}[ht]
\begin{center}
\includegraphics[width=8cm]{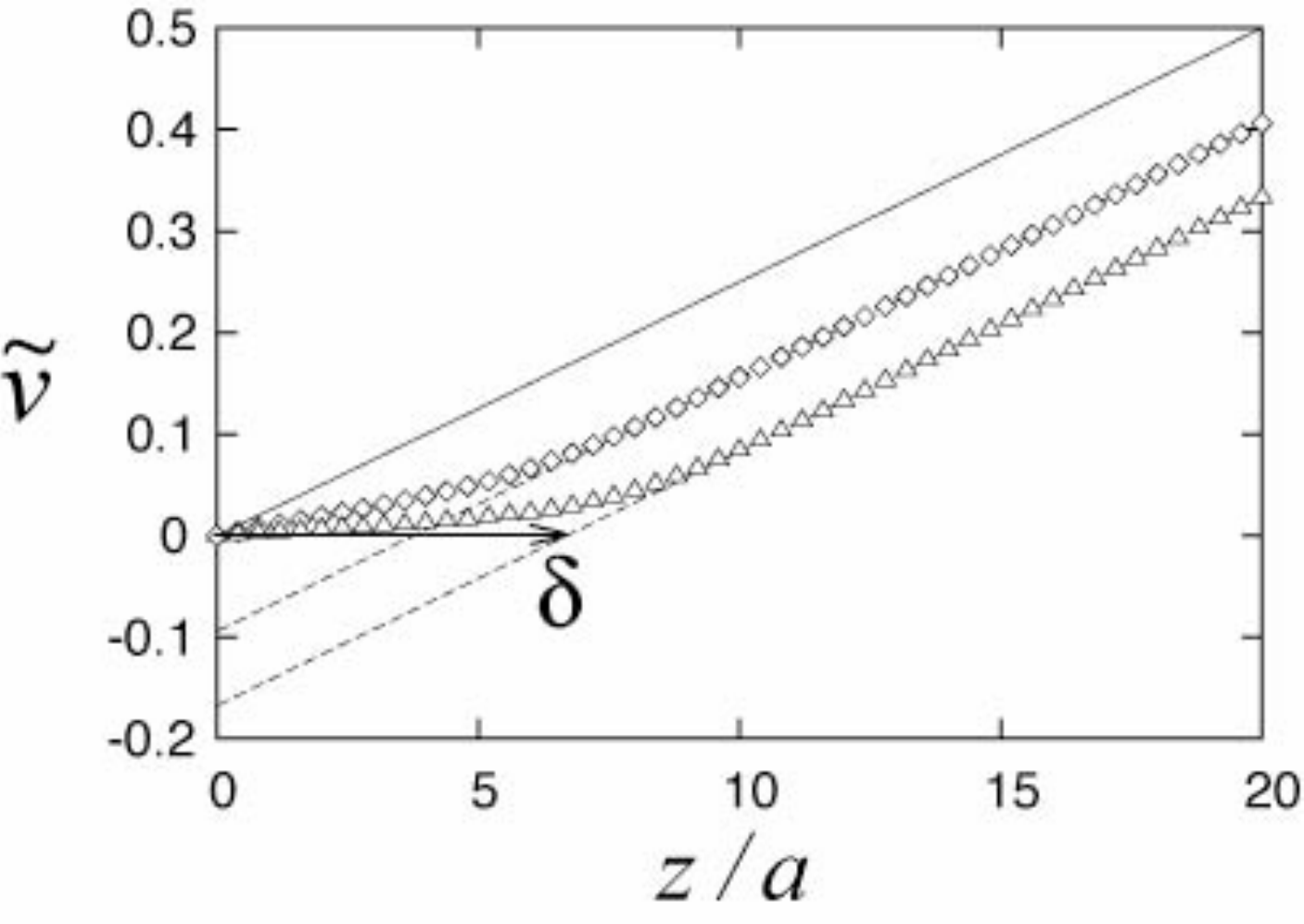}
\caption{Solvent flow profiles  as a function of distance from the surface
for grafting densities $\rho=0.003$ (diamonds) and $\rho=0.007$ (triangles).
Parameters of the system are chosen as $\ell_p/L=5$ and $\dot{\tilde{\gamma}} =0.03$
(which corresponds to $\dot{\gamma} \approx 10^8$ s$^{-1}$ for $a=1$ nm).
The length of polymers is roughly $L = 20  a$. The effective shear-dependent stagnation length $\delta$ is determined by a linear extrapolation of the flow profile towards the grafting surface (dashed lines). For comparison, the unperturbed flow velocity in the absence of grafted chains is denoted by a solid line.}
\label{fig10}
\end{center}
\end{figure}

Figure~\ref{fig10} shows the stationary solvent velocity profiles $ \tilde{v} (\tilde{z}) = \Delta v/(12 \pi \tilde{\dot{\gamma}} L)$
measured from the simulation, and clearly exhibits a 
shear-dependent lift-up of the shear plane. 
By linear extrapolation of the flow profile from large distances to the
surface, the shear-dependent stagnation thickness $\delta$
(corresponding to the lift-up of the shear plane)
in units of $a$ is quantitatively determined in the figure (dashed lines).
It is numerically difficult to reach the small shear rates typically  used in experiments.
We instead perform simulations for
elevated shear rates at which numerical errors are small and extrapolate down to 
small shear rates. Let us discuss the experimental relevance of parameters used in the simulation:
we rescaled every parameter using monomer radius $a$ and thermal energy $k_BT$.
Therefore, choosing a proper value of $a$, we are able to cover
a wide range of experimental situations.
For example, when $a=1$ nm is assumed, as applicable to DNA,
 it gives a contour length $L=20$ nm and shear rate
$\dot{\gamma} \approx10^8$ s$^{-1}$ for $\dot{\tilde{\gamma}}  =0.03$ (which is a quite big
number in a conventional set-up).
However, assuming $a=100$ nm, one gets $L=2 \mu$m and
$\dot{\gamma} \approx10^2$ s$^{-1}$ for $\dot{\tilde{\gamma}}   =0.03$, which
can be easily reached in the laboratory.

\begin{figure}[ht]
\begin{center}
\includegraphics[width=8cm]{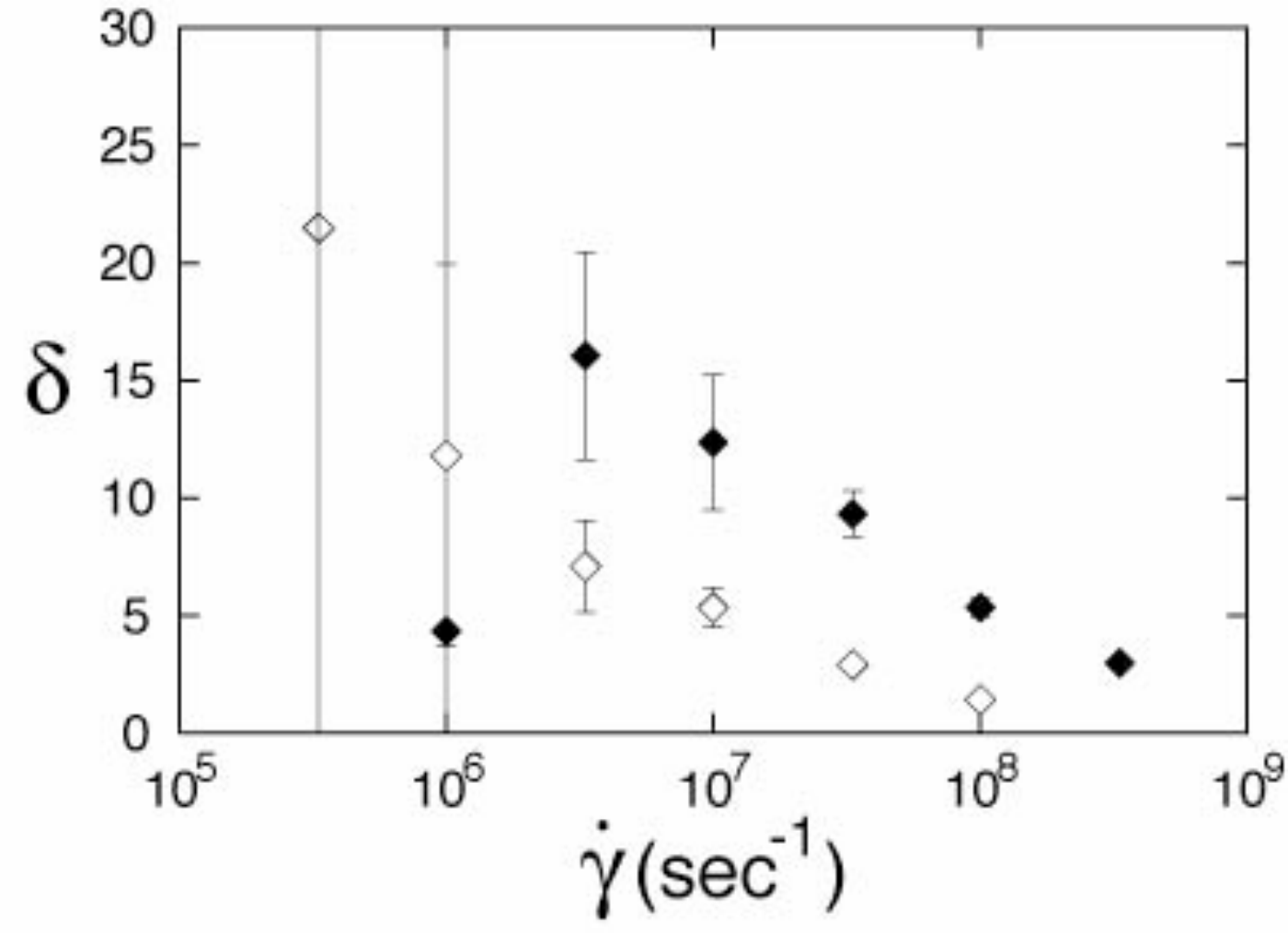}
\caption{Stagnation length $\delta$ vs. shear rate $\dot{\gamma}$ for two different grafting densities: $\rho=0.005$ (filled diamonds) and 0.001 (open diamonds).The unitless shear rate $\tilde{ \dot{\gamma}}$ actually used in the simulation has been converted into an explicit shear rate $\dot{\gamma} $ \textit{via} $\dot{\gamma}=k_BT \tilde { \dot{\gamma}} /a^3 \eta$ and assuming a monomer radius of $a=1$ nm, as applicable to DNA. The persistence of the filament is $\ell_p/L=5$.}
\label{fig11}
\end{center}
\end{figure}

In order to compare with the experimental results, we study the stagnation thickness for
different applied shear rates. As shown in Fig.~\ref{fig11}, the stagnation length $\delta$ decreases for increasing shear rate. While the chains are strongly bent close to the surface under strong shear, they tend to keep an upright conformation under weak shearing, leading to significantly enhanced hydrodynamic friction.

\begin{figure}[ht]
\begin{center}
\includegraphics[width=8cm]{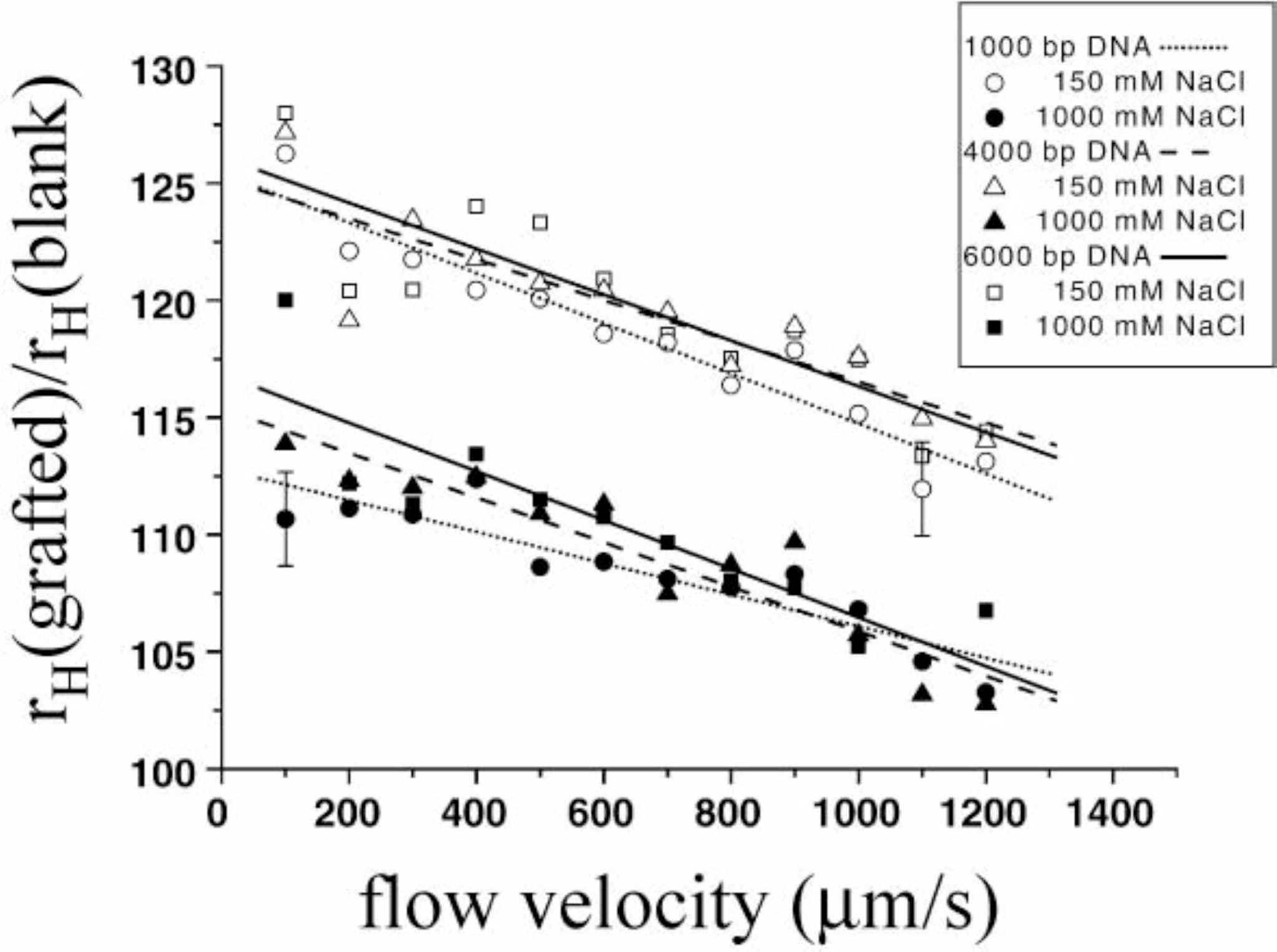}
\caption{Ratio of the effective hydrodynamics radii $r_H$ for DNA-grafted and blank colloids vs. flow velocity. Reproduced from ~\cite{Gutsche} .
The symbols corresponds to three colloids with different DNA lengths ($\circ$: 1000 bp; $\triangle$: 4000 bp; $\square$: 6000 bp and grafting density of $0.03 \pm 0.01 \mu m^2$) in two NaCl electrolytes (open symbols: 150 mM; filled symbols: 100 mM.
The results show that the hydrodynamic radius decreases with increasing shear rate,
indicative of  conformational changes of the grafted polymers and an effective
decrease of the brush layer.}
\label{gutsche}
\end{center}
\end{figure}

This is in qualitative agreement with recent laser-trap experiments on DNA chains that
are anchored on spherical colloids~\cite{Gutsche} which results are reproduced in Fig.~\ref{gutsche}. 
In the experiments the brush-covered colloids, which have a diameter
of roughly a micrometer, are held fixed in the laser trap and subjected
to laminar flows of the order of up to 1000 $\mu m$/s, giving rise to shear rates
of the order of $1000$~s$^{-1}$. The force acting on the colloid can be measured
and thus the hydrodynamic radius in inferred.
We observe that the hydrodynamic radius of DNA-grafted colloids decreases with the flow velocity, i.e. the shear rate in our model. 
Hence, the stagnation length decrease with increasing shear rate, 
which constitutes a non-linear friction response of the suspended colloidal sphere as a function of solvent velocity.  This is in qualitative agreement with the simulations.

\section{Conclusion}

Three different situations have been reviewed having in common that the coupling between 
hydrodynamic interactions and elastic deformations of soft matter plays an important role.
When a significant external force is applied to the monomers of a deformable polymer, 
a complex hydrodynamic interaction mediated by the surrounding water sets in and 
couples the velocities of  different components of the system. 
This can lead to unexpected motions of the whole system: \textit{i)} a sedimenting elastic rod deforms and, as a consequence, aligns perpendicularly to the direction of motion;
\textit{ii)} a rotating filament deforms and therefore gives rise to propulsion perpendicularly to 
the rotation axis, and \textit{iii)} the deformation of surface anchored polymers in shear flow leads to a shear-rate dependent shift of the stagnation-plane position.
Since all materials have actually finite elastic moduli, their deformations in flow fields are unavoidable. These phenomena are therefore relevant to recent experiments
probing the coupling of elasticity and hydrodynamics of nano-systems far from equilibrium.
It is suggested that hydrodynamic interactions should also be taken into account in 
biological systems~\cite{camalet}. Indeed, as an example, one may consider the propulsion mechanism studied in Section~\ref{prop} and carry the lesson to bacterial propulsion. 
Of course, flagella are helical and their physics is therefore much more involved than a simple straight
propeller. However, flagellar motors generate sufficient torques for the non-linear elastic phenomena discussed in this review to occur: they are powered by a proton-motive force which yields torques on the order of $N \approx 10^3\,\,k_BT$~\cite{berry,turner}. Moreover, the flagellum length is $L\simeq 10\;\mu$m, leading to $\ell_{\mathrm{p}}/L\simeq 10^3-10^4$ and $\tilde{N}L/\ell_{\mathrm{p}}\simeq 0.1-1$. 
Hence, torque-induced shape transformations are most likely biologically relevant and might be directly observed with straight biopolymers attached to flagellar motors. Indeed, 
 the role of flexibility in bacterial propulsion  has been recently stressed since bending is crucial for the bundling of flagella~\cite{kim,Powers}.\\

This work was financially supported by Deutsche Forschungsgemeinschaft
(DFG, German-French Network, SPP 1164 Nano- and Microfluidics, and SFB563)
and the Fonds der Chemischen Industrie.
YWK also acknowledges support from NSF DMR 0080034.

\end{document}